# Thermodynamics of mixtures with strongly negative deviations from Raoult's law. XVI. Permittivities and refractive indices for 1-alkanol + di-*n*-propylamine systems at (293.15-303.15) K. Application of the Kirkwood-Fröhlich model

Fernando Hevia, Ana Cobos, Juan Antonio González*, Isaías García de la Fuente, Luis Felipe Sanz

G.E.T.E.F., Departamento de Física Aplicada, Facultad de Ciencias, Universidad de Valladolid. Paseo de Belén, 7, 47011 Valladolid, Spain.

*e-mail: jagl@termo.uva.es; Tel: +34-983-423757

# Abstract

Relative permittivities at 1 MHz, $\varepsilon_r$, and refractive indices at the sodium D-line, $n_D$, are reported at 0.1 MPa and at (293.15-303.15) K for the binary systems 1-alkanol + di-*n*-propylamine (DPA). Their corresponding excess functions are calculated and correlated. For the methanol mixture, positive values of the excess permittivities, $\varepsilon_r^E$, are found. Except at high concentrations of the alcohol in the 1-propanol mixture, the remaining systems show negative values of this property. This fact reveals that the creation of (1-alkanol)-DPA interactions contributes positively to $\varepsilon_r^E$, being this contribution dominant in the methanol mixture. The negative contributions arising from the disruption of interactions between like molecules are prevalent in the other mixtures. At $\phi_1$ (volume fraction) = 0.5, $\varepsilon_r^E$ changes in the sequence: methanol > 1-propanol > 1-butanol > 1-pentanol < 1-heptanol. An analogous variation with the chain length of the 1-alkanol is observed in mixtures such as 1-alkanol + heptane, + cyclohexylamine or + *n*-hexylamine (HxA). Moreover, for a given 1-alkanol, $\varepsilon_r^E$ is larger for DPA than for HxA mixtures, suggesting that in DPA solutions multimers with parallel alignment of the molecular dipoles are favoured and cyclic multimers are disfavoured when compared to HxA mixtures. The $(\partial \varepsilon_r / \partial T)_p$ values are higher for the mixtures than for pure 1-alkanols, because (1-alkanol)-DPA interactions are stronger than those between 1-alkanol molecules. Calculations on molar refractions indicate that dispersive interactions in the systems under study increase with the chain length of the 1-alkanol and are practically identical to those in HxA solutions. The considered mixtures are treated by means of the Kirkwood-Fröhlich model, reporting the Kirkwood correlation factors and their excess values.

Keywords: 1-alkanol; di-*n*-propylamine; permittivity; refractive index; Kirkwood correlation factor.

# 1. Introduction

Amines are found in situations of biological interest. For instance, the breaking of amino acids releases amines and proteins that are usually bound to DNA polymers contain several amine groups [1]. Their low vapour pressure makes them useful in green chemistry. Mixtures containing amines are being investigated to be used in $CO_2$ capture [2]. On the other hand, many of the ions of the technically important ionic liquids are related to amine groups [3]. Linear primary and secondary amines are weakly self-associated compounds [4-8] with rather low dipole moments. Liquid mixtures formed by 1-alkanol and a linear primary or secondary amine are rather interesting from a theoretical point of view, as they show strongly negative deviations from Raoult's law. In fact, the excess molar Gibbs energies, $G_m^E$, at $x_1$ (mole fraction) = 0.5 for methanol systems are: –823 J·mol$^{-1}$ (di-$n$-ethylamine; $T$ = 298.15 K [9]) and –799 J·mol$^{-1}$ ($n$-butylamine, $T$ = 348.15 K [10]). Accordingly, the excess molar enthalpies ($H_m^E$) are large and negative. For instance, at 298.15 K and $x_1 = 0.5$; $H_m^E$ (methanol)/J·mol$^{-1}$ = –3200 ($n$-hexylamine (HxA)) [11]; –4581 (di-$n$-ethylamine) [12]. This has been explained in terms of two different opposing effects. In the pure liquid state, both 1-alkanols and linear amines are self-associated by means of O-H---O and N-H---N bonds, respectively. Such bonds are disrupted along the mixing process, which positively contribute to $H_m^E$. On the other hand, it is well known that the formation of interactions between unlike molecules upon mixing contribute negatively to $H_m^E$. Therefore, the large and negative $H_m^E$ values of this type of systems reveal that the new O-H---N bonds created are stronger than the O-H---O and N-H---N bonds. For instance, the values of the enthalpy of the hydrogen bonds between methanol and amine estimated from the application of the ERAS model [13] are: –42.4 kJ·mol$^{-1}$ ($n$-hexylamine) [7]; –45.4 kJ·mol$^{-1}$ (di-$n$-ethylamine) [14]. We remark that such values are much more negative than that used, within this model, for the enthalpy of the H bonds between alkanol molecules, –25.1 kJ·mol$^{-1}$ [7, 13, 14]. As a consequence of the strong interactions between unlike molecules, the systems are highly structured. For example, at $T$ = 298.15 K and $x_1 = 0.5$, $TS_m^E$ ($= H_m^E - G_m^E$) is –3758 J·mol$^{-1}$ for the methanol + di-$n$-ethylamine mixture (see above). This result is much more negative than the value for the 1-propanol + hexane system, $TS_m^E = (533 (= H_m^E) - 1295 (= G_m^E)) = -762$ J·mol$^{-1}$ [15, 16]. The large and negative excess molar volumes [7, 17-21] and solid-liquid equilibria (SLE) measurements [22] also support the existence of strong interactions between unlike molecules in 1-alkanol + linear amine mixtures. It is to be noted that the SLE phase diagrams show that complex formation is an important feature of these solutions [22]. In addition, $\varepsilon_r^E$ values also indicate strong interactions between unlike molecules in 1-alkanol +

linear primary amine systems; e.g. for the methanol + HxA mixture [23] $\varepsilon_r^E$ = 1.480 at $T$ = 298.15 K and $\phi_1$ (volume fraction) = 0.5.

We have extended the database of 1-alkanol + amine mixtures reporting excess molar volumes [7, 17-21]; dynamic viscosities [19-21]; vapour-liquid equilibria [24]; permittivities ($\varepsilon_r$) and refractive indices ($n_D$) [19-21, 25]. In addition, these systems have been investigated using different models as DISQUAC or ERAS [6, 7, 14, 17, 18, 20, 26-28]; the formalism of the Kirkwood-Buff integrals [29], or the concentration-concentration structure factor ($S_{CC}(0)$) formalism [30]. More recently [23], we have provided $\varepsilon_r$ and $n_D$ data for the 1-alkanol + HxA mixtures over the temperature range (293.15-303.15) K, and analysed them using the Kirkwood-Fröhlich model [31-34], which is a useful approach to gain insight into the structure and interactions of mixtures. As a continuation, and in order to investigate the effect of replacing a linear primary amine (HxA) by a linear secondary amine (di-*n*-propylamine, (DPA)), we report similar measurements over the same range of temperature for mixtures formed by the latter amine and methanol, or 1-propanol, or 1-butanol, or 1-pentanol or 1-heptanol. In addition, the systems are also studied by means of Kirkwood-Fröhlich model.

## 2. Experimental

*2.1 Materials*

Information about the purity and source of the pure compounds, which were used in the experiments without further purification, is collected in Table 1. Their $\varepsilon_r$ values at 1 MHz, densities ($\rho$) and $n_D$ values at 0.1 MPa and at the working temperatures can be found in Table 2. These results agree well with literature data.

*2.2 Apparatus and procedure*

Binary mixtures were prepared by mass in small vessels of about 10 cm$^3$ with the aid of an analytical balance Sartorius MSU125p (weighing accuracy 0.01 mg), taking into account the corresponding corrections on buoyancy effects. The standard uncertainty in the final mole fraction is 0.0010. Molar quantities were calculated using the relative atomic mass Table of 2015 issued by the Commission on Isotopic Abundances and Atomic Weights (IUPAC) [35]. In order to minimize the effects of the interaction of the compounds with air components, they were stored with 4 Å molecular sieves (except methanol, because measurements were affected). In addition, the measurement cell (see below) was completely filled with the samples and appropriately closed. Different density measurements of pure compounds, conducted along experiments, showed that this quantity remained unchanged within the experimental uncertainty.

Temperatures were measured with Pt-100 resistances, calibrated according to the ITS-90 scale of temperature, against the triple point of water and the fusion point of Ga. The standard uncertainty of this quantity is 0.01 K for $\rho$ determinations, and 0.02 K for $\varepsilon_r$ and $n_D$ measurements.

The $\varepsilon_r$ measurements were performed with the aid of an equipment from Agilent. A 16452A cell, which is a parallel-plate capacitor made of Nickel-plated cobalt (54% Fe, 17% Co, 29% Ni) with a ceramic insulator (alumina, $Al_2O_3$), is filled with a sample volume of $\approx 4.8$ cm$^3$. The cell is connected by a 16048G test lead to a precision impedance analyser 4294A, and immersed in a thermostatic bath LAUDA RE304, with a temperature stability of 0.02 K. Details about the device configuration and calibration are given elsewhere [36]. The relative standard uncertainty of the $\varepsilon_r$ measurements (i.e. the repeatability) is 0.0001. The total relative standard uncertainty of $\varepsilon_r$ was estimated to be 0.003 from the differences between our data and values available in the literature, in the range of temperature (288.15‑333.15) K, for the following pure liquids: water, benzene, cyclohexane, hexane, nonane, decane, dimethyl carbonate, diethyl carbonate, methanol, 1-propanol, 1-pentanol, 1-hexanol, 1-heptanol, 1-octanol, 1-nonanol and 1-decanol.

A Bellingham+Stanley RFM970 refractometer was used for the $n_D$ measurements. The technique is based on the optical detection of the critical angle at the wavelength of the sodium D line (589.3 nm). The temperature is controlled by Peltier modules and its stability is 0.02 K. The refractometer has been calibrated using 2,2,4-trimethylpentane and toluene at (293.15-303.15) K, following the recommendations by Marsh [37]. The standard uncertainty of $n_D$ is 0.00008.

Densities were obtained using a vibrating-tube densimeter and sound analyzer Anton Paar DSA5000, which is automatically thermostated within 0.01 K. The calibration procedure has been described elsewhere [38]. The relative standard uncertainty of the $\rho$ measurements is 0.0012.

## 3. Results

Let us denote by $x_i$ the mole fraction of component $i$. The corresponding volume fraction, $\phi_i$, is given by $\phi_i = x_i V_{mi}^* / \left( x_1 V_{m1}^* + x_2 V_{m2}^* \right)$, where $V_{mi}^*$ stands for the molar volume of component $i$. For an ideal mixture at the same temperature and pressure as the mixture under study, the relative permittivity, $\varepsilon_r^{id}$, the derivative $\left[\left(\partial \varepsilon_r / \partial T\right)_p\right]^{id}$, and the refractive index, $n_D^{id}$, are given by [39, 40]:

$$\varepsilon_r^{id} = \phi_1 \varepsilon_{r1}^* + \phi_2 \varepsilon_{r2}^* \tag{1}$$

$$n_D^{id} = \left[\phi_1 \left(n_{D1}^*\right)^2 + \phi_2 \left(n_{D2}^*\right)^2\right]^{1/2} \tag{2}$$

$$\left[\left(\frac{\partial \varepsilon_r}{\partial T}\right)_p\right]^{id} = \left(\frac{\partial \varepsilon_r^{id}}{\partial T}\right)_p \tag{3}$$

where $\varepsilon_{ri}^*$ and $n_{Di}^*$ denote the relative permittivity and the refractive index of pure species $i$, and $\left(\partial \varepsilon_r^{id} / \partial T\right)_p$ is calculated from linear regressions as indicated below. The corresponding excess functions, $F^E$, are obtained as

$$F^E = F - F^{id} \quad , \quad F = \varepsilon_r, \, n_D, \, \left(\frac{\partial \varepsilon_r}{\partial T}\right)_p \tag{4}$$

Table 3 lists $\phi_1$, $\varepsilon_r$ and $\varepsilon_r^E$ values of 1-alkanol (1) + DPA (2) systems as functions of $x_1$, in the temperature range (293.15 – 303.15) K. Table 4 contains the experimental $x_1$, $\phi_1$, $n_D$ and $n_D^E$ values.

We calculated the derivative $\left(\partial \varepsilon_r / \partial T\right)_p$ at 298.15 K as the slope of a linear regression of experimental $\varepsilon_r$ values in the range (293.15 – 303.15) K. The data of $\left[\left(\partial \varepsilon_r / \partial T\right)_p\right]^E = \left(\partial \varepsilon_r^E / \partial T\right)_p$ are collected in Table S1 (supplementary material).

The $F^E$ data were fitted to a Redlich-Kister equation [41] by unweighted linear least-squares regressions:

$$F^E = x_1 (1 - x_1) \sum_{i=0}^{k-1} A_i (2x_1 - 1)^i \tag{5}$$

The number, $k$, of necessary coefficients for this regression has been determined, for each system and temperature, by applying an F-test of additional term [42] at a 99.5% confidence level. Table 5 includes the parameters $A_i$ obtained, and the standard deviations $\sigma(F^E)$, defined by:

$$\sigma(F^E) = \left[\frac{1}{N-k} \sum_{j=1}^{N} \left(F_{cal,j}^E - F_{exp,j}^E\right)^2\right]^{1/2} \tag{6}$$

where the index $j$ takes one value for each of the $N$ experimental data $F^E_{\exp,j}$, and $F^E_{\text{cal},j}$ is the corresponding value of the excess property $F^E$ calculated from equation (5).

Values of $\varepsilon_r^E$, $n_D^E$ and $\left(\partial \varepsilon_r^E / \partial T\right)_p$ versus $\phi_1$ of 1-alkanol + DPA systems at 298.15 K are plotted in Figures 1, 2 and 3 respectively with their corresponding Redlich-Kister regressions. Data on $n_D$ are plotted in Figure S1 (supplementary material).

## 4. Discussion

Unless stated otherwise, the below values of the thermophysical properties and their corresponding excess functions are referred to $T = 298.15$ K and $\phi_1 = 0.5$. We will denote by $n$ the number of C atoms of the 1-alkanol.

### 4.1. Excess relative permittivities

It is known that the breaking of interactions between like molecules in the mixing process leads to a negative contribution to $\varepsilon_r^E$. On the other hand, the creation of interactions between molecules of different species can lead to either a positive or to a negative contribution to $\varepsilon_r^E$, depending on the capability of the multimers formed to respond to an external electric field and lead to a macroscopic dipole moment. For instance, 1-alkanol + heptane mixtures show rather large and negative values of this quantity, which can be ascribed to the breaking of the 1-alkanol self-association (Figure 4): $\varepsilon_r^E = -1.075$ ($n = 3$), $-2.225$ ($n = 4$), $-2.525$ ($n = 5$), $-2.875$ ($n = 7$), $-1.775$ ($n = 10$) [20, 43-45]. For methanol, there exists a partial immiscibility region [46]. The corresponding $\varepsilon_r^E$ values of 1-alkanol + DPA systems are higher: 2.406 ($n = 1$), $-0.246$ ($n = 3$), $-0.715$ ($n = 4$), $-0.883$ ($n = 5$), $-0.747$ ($n = 7$) (Figure 4). This reveals that alkanol-amine interactions contribute positively to the polarization of the mixture. The positive $\varepsilon_r^E$ result for the methanol + DPA system strongly confirms this conclusion. 1-Alkanol + cyclohexylamine [21, 25], or + HxA [23] mixtures behave similarly and also show higher $\varepsilon_r^E$ values than those of 1-alkanol + heptane systems (Figure 4). On the other hand, the $\varepsilon_r^E(n)$ variation for 1-alkanol + DPA, or + cyclohexylamine, or + HxA mixtures follows the sequence: methanol > 1-propanol > 1-butanol > 1-pentanol < 1-heptanol (Figure 4), which is similar to that encountered for 1-alkanol + heptane mixtures (see above, Figure 4). For the latter systems, it has been explained in terms of the lower and weaker self-association of longer 1-alkanols [25]. For amine systems, this statement is still valid, but interactions between unlike molecules must be also considered. Studies on 1-alkanol + amine mixtures using the ERAS model show that solvation effects between unlike molecules decrease when the alkanol size is increased [7, 17, 18]. This means that the polarization changes, along the mixing process, to a lower extent when longer 1-alkanols are involved, since these alcohols are less self-associated and the corresponding solvation effects are also less important. It is to be noted that $\varepsilon_r^E$ changes more sharply when increasing $n$ for mixtures with shorter 1-alkanols than for systems involving longer 1-alkanols and that the same occurs for the excess molar volumes and for the excess molar enthalpies [17].

Interestingly, for a given 1-alkanol, say 1-butanol, $\varepsilon_r^E$(DPA) = –0.715 > $\varepsilon_r^E$(HxA) = –1.424 [15] (Figure 4). This suggests that in DPA solutions multimers with parallel alignment of the molecular dipoles are favoured and cyclic multimers are disfavoured when compared to HxA mixtures. Furthermore, at $\phi_1 = 0.47$, the 1-butanol + di-$n$-ethylamine mixture [47] shows an even higher value (–0.13), which can be explained by the formation of more and stronger H bonds between unlike molecules, because the amine group is less sterically hindered in this amine.

It may be pertinent to compare the dielectric behaviour of mixtures formed by 1-alkanol and DPA or di-$n$-propylether (DPE), as both solvents have similar size and structure. It is well known that the thermodynamic properties of the DPE systems are mainly characterized by the alkanol self-association [48]. Thus, the $H_m^E$ values are moderately positive ($H_m^E$/J·mol$^{-1}$ = 740 for the 1-propanol system [49]); remain nearly constant for mixtures involving the longer 1-alkanols, and the corresponding $H_m^E$ curves are shifted towards low mole fractions of the 1-alkanol [48]. In contrast, as it has been previously mentioned, solvation, i.e. strong interactions between unlike molecules, is the main feature of 1-alkanol + DPA mixtures [14]. This is clearly demonstrated by the large and negative $H_m^E$ values of these systems (– 2432 J·mol$^{-1}$ for the 1-butanol solution [50]). For DPE mixtures, the dependence of $\varepsilon_r^E$ with the alcohol size is similar to that encountered for the amine systems examined: –1.03 (ethanol) < –1.24 (1-butanol) < –1.60 (1-hexanol) > –0.80 (1-decanol) [51]. On the other hand, for mixtures with a given 1-alkanol, $\varepsilon_r^E$ changes in the order: heptane < DPE < DPA (see above, Figure 4). This reveals that interactions between unlike molecules contribute more positively to the polarization of the mixture in the case of DPA systems.

### 4.2. Molar refraction

The refractive index at optical wavelengths is closely related to dispersion forces, since the molar refraction (or molar refractivity), $R_m$ [34, 52]:

$$R_m = \frac{n_D^2 - 1}{n_D^2 + 2} V_m = \frac{N_A \alpha_e}{3\varepsilon_0} \qquad (7)$$

(where $N_A$ and $\varepsilon_0$ stand for Avogadro's constant and the vacuum permittivity, respectively) is proportional to the mean electronic polarizability, $\alpha_e$ [32, 34]. For the investigated systems, the values of $R_m$ / cm$^3$·mol$^{-1}$ at $x_1 = 0.5$ are (Figure S2, supplementary material): 20.5 ($n = 1$), 25.2 ($n = 3$), 27.5 ($n = 4$), 29.8 ($n = 5$), 34.4 ($n = 7$). It is clear that dispersive interactions are more important in longer 1-alkanols. Moreover, the values are practically identical to those of 1-alkanol + HxA mixtures [23]. This is to be expected, as DPA and HxA are isomers and both linear, so dispersive interactions cannot differ appreciably. The excess molar refractions, $R_m^E = R_m - R_m^{id}$, have also been calculated, with $R_m^{id}$ evaluated substituting ideal values in equation (7). Values of $R_m^E$ for 1-alkanol + hexane ($n = 3, 4, 5, 6, 8$ [53, 54]) are positive and small (< 0.04 cm$^3$·mol$^{-1}$). The same occurs for DPA + heptane (< 0.07 cm$^3$·mol$^{-1}$ [45, 55], assuming ideal behaviour of $n_D$). However, in 1-alkanol + DPA systems the curves are negative; at $x_1 = 0.5$, $R_m^E$ / cm$^3$·mol$^{-1}$ –0.45 ($n = 1$), –0.37 ($n = 3$), –0.38 ($n = 4$), –0.39 ($n = 5$), –0.40 ($n = 7$). This loss in dispersive interactions along mixing with respect to the ideal state can then be ascribed to a large number of O-H---N bonds formed in the mixing process, being greater for the methanol mixture.

### 4.3. Temperature dependence of the permittivity

Firstly, we note that, for pure compounds, $(\partial \varepsilon_r^* / \partial T)_p$ values are negative (Table 6), which is the typical behaviour of normal liquids. In the case of 1-alkanols, this quantity increases with $n$ since the alcohol self-association decreases at this condition and a lower number of interactions between alcohol molecules are broken when the temperature is increased. The higher $(\partial \varepsilon_r^* / \partial T)_p$ values of DPA or HxA can be explained similarly. Interestingly, results for $(\partial \varepsilon_r / \partial T)_p$ are larger for the considered systems than for pure 1-alkanols (Table 6), which underlines the existence of (1-alkanol)-amine interactions. It is known that such interactions are stronger than those between alcohol molecules. For example, in the framework of the ERAS model, as already mentioned, the enthalpy of the hydrogen bonds between 1-alkanol molecules is –25 kJ·mol$^{-1}$ [7, 11, 13, 17] while the enthalpies between methanol or 1-heptanol and DPA molecules are, respectively, –42.4 and –34.5 kJ·mol$^{-1}$ [17]. Thus, one can expect that the number of (1-alkanol)-amine interactions broken when the temperature is increased is lower than the number of disrupted interactions between 1-alkanol molecules. This makes $\varepsilon_r$ change more smoothly with temperature for the mixtures than for pure 1-alkanols since, as it has been previously indicated, (1-alkanol)-amine interactions contribute positively to the polarization of the system. On the other hand, $(\partial \varepsilon_r / \partial T)_p$ also increases in line with $n$. The weaker temperature dependence of $\varepsilon_r$ for systems containing longer 1-alkanols can be newly explained as above, i.e., in terms of the lower self-association of these 1-alkanols and of the less important solvation effects involved. We also note that $(\partial \varepsilon_r^E / \partial T)_p$ may show either positive or negative values (Table 5). Negative values (systems with $n$ = 1-4) mean that $\varepsilon_r$ decreases with the increase of temperature more rapidly than $\varepsilon_r^{id}$ does. This behaviour is encountered for solutions where the effects related to the alcohol self-association and solvation effects between unlike molecules are more relevant. They become less important in systems with $n$ = 5,7, and the temperature dependence of $\varepsilon_r$ is weaker than that of $\varepsilon_r^{id}$, leading to positive $(\partial \varepsilon_r^E / \partial T)_p$ values. Finally, the replacement of DPA by HxA in systems with a given 1-alkanol leads to less negative $(\partial \varepsilon_r / \partial T)_p$ values (Table 6). This newly suggests that cyclic multimers formed by unlike molecules also exist in 1-alkanol + HxA systems, as the disruption of such multimers for increased temperature values positively contributes to the mixture polarization.

### 4.4. Kirkwood-Fröhlich model

In the Kirkwood-Fröhlich model, the fluctuations of the dipole moment in the absence of the electric field are treated as the basis to obtain relations involving the relative permittivity. It is a local-field model in which the molecules are assumed to be in a spherical cavity and the induced contribution to the polarizability is treated macroscopically through its relation to $\varepsilon_r^\infty$ (the value of the permittivity at a high frequency at which only the induced polarizability contributes). The local field takes into account long-range dipolar interactions by considering the outside of the cavity as a continuous medium of permittivity $\varepsilon_r$. Short-range interactions are introduced by the so-called Kirkwood correlation factor, $g_K$, which provides information about the deviations from randomness of the orientation of a dipole with respect to its neighbours. This is an important parameter, as it provides information about specific interactions in the liquid state. For a mixture, $g_K$ can be determined, in the context of a one-fluid model [31], from macroscopic physical properties according to the expression [31-34]:

$$g_K = \frac{9k_B T V_m \varepsilon_0 (\varepsilon_r - \varepsilon_r^\infty)(2\varepsilon_r + \varepsilon_r^\infty)}{N_A \mu^2 \varepsilon_r (\varepsilon_r^\infty + 2)^2} \qquad (8)$$

Here, $k_B$ is Boltzmann's constant; $N_A$, Avogadro's constant; $\varepsilon_0$, the vacuum permittivity; and $V_m$, the molar volume of the liquid at the working temperature, $T$. For polar compounds, $\varepsilon_r^\infty$ is estimated from the relation $\varepsilon_r^\infty = 1.1 n_D^2$ [56]. $\mu$ represents the dipole moment of the solution, estimated from the equation [31]:

$$\mu^2 = x_1\mu_1^2 + x_2\mu_2^2 \qquad (9)$$

where $\mu_i$ stands for the dipole moment of component i (=1,2). Calculations have been performed using smoothed values of $V_m^E$ [17], $n_D^E$ (this work) and $\varepsilon_r^E$ (this work) at $\Delta x_1 = 0.01$. The source and values of $\mu_i$ are collected in Table 2.

Figure 5 shows our calculations on $g_K$ of 1-alkanol + DPA systems, which takes the values: 2.97 (n = 1), 2.72 (n = 3), 2.60 (n = 4), 2.47 (n = 5), 2.25 (n = 7). These are greater than the corresponding values for 1-alkanol + HxA mixtures [23]: 2.68 (n = 1), 2.32 (n = 3), 2.16 (n = 4), 2.01 (n = 5), 1.77 (n = 7). This would mean that parallel alignment of the dipoles is more favoured in DPA mixtures, supporting our previous statement inferred from the analysis of $\varepsilon_r^E$. It is interesting to note that for $\phi_1 > 0.4$ the $g_K$ curve for methanol + DPA is practically constant, suggesting that the structure of the mixture in this concentration range is quite similar to that of the pure methanol because the rupture of the methanol self-association is compensated by the methanol-DPA hydrogen bonds created.

We have calculated as well the excess Kirkwood correlation factors, $g_K^E = g_K - g_K^{id}$, where $g_K^{id}$ is calculated substituting the real quantities by ideal ones in equation (8). The values for 1-alkanol + DPA systems are (Figures 6 and 7): 0.317 (n = 1), –0.110 (n = 3), –0.270 (n = 4), –0.366 (n = 5), –0.377 (n = 7). The trend is parallel to that of 1-alkanol + HxA mixtures [23], being this, as the corresponding $\varepsilon_r^E$, lower (Figure 7): 0.170 (n = 1), –0.257 (n = 3), –0.421 (n = 4), –0.505 (n = 5), –0.508 (n = 7). The interpretation of this fact is thus similar [23]. For the minimum of the curves, the variation is the same as the one encountered for $\varepsilon_r^E$, but it occurs at lower values of $\phi_1$. Then, according to the model, the $\varepsilon_r^E$ minima are influenced by other factors different from the variation of the correlations in the orientation of the dipoles in the mixing process.

## 5. Conclusions

Measurements of $\varepsilon_r$ and $n_D$ have been reported for the 1-alkanol + di-n-propylamine systems at (293.15-303.15) K. Interactions between unlike molecules form multimers that contribute positively to $\varepsilon_r^E$. Such contribution is dominant for the methanol mixture and $\varepsilon_r^E$ is positive. For the remaining systems (except for high $\phi_1$ values in the 1-propanol mixture) $\varepsilon_r^E$ values are negative, indicating that dominant contributions arise from the breaking of interactions between like molecules. For a given 1-alkanol, $\varepsilon_r^E$ is larger for di-n-propylamine than for n-hexylamine mixtures. This suggests that parallel alignment of the dipoles is more favoured and cyclic multimers disfavoured in the former case. The behaviour of $(\partial \varepsilon_r / \partial T)_p$ and the application of the Kirkwood-Fröhlich model support these findings. The values of $(\partial \varepsilon_r / \partial T)_p$ are higher for the mixtures than for pure 1-alkanols, because (1-alkanol)-DPA interactions are stronger than those between 1-alkanol molecules. Calculations on $R_m$ show that dispersive interactions in the studied mixtures increase with the length of the 1-alkanol, and that they have the same importance as in *n*-hexylamine systems.

## Acknowledgements

F. Hevia and A. Cobos are grateful to Ministerio de Educación, Cultura y Deporte for the grants FPU14/04104 and FPU15/05456 respectively. The authors gratefully acknowledge the financial support received from the Consejería de Educación, Junta de Castilla y León, under Project BU034U16.

Table 1

Sample description.

| Chemical name | CAS Number | Source | Purification method | Purity[a] |
|---|---|---|---|---|
| methanol | 67-56-1 | Sigma-Aldrich | none | 99.99% |
| 1-propanol | 71-23-8 | Sigma-Aldrich | none | 99.84% |
| 1-butanol | 71-36-3 | Sigma-Aldrich | none | 99.86% |
| 1-pentanol | 71-41-0 | Sigma-Aldrich | none | 99.9% |
| 1-heptanol | 111-70-6 | Sigma-Aldrich | none | 99.9% |
| di-*n*-propylamine (DPA) | 111-26-2 | Aldrich | none | 99.9% |

[a] In mole fraction. By gas chromatography. Provided by the supplier.

Table 2

Dipole moment, $\mu$, of the pure compounds, and their relative permittivity at frequency $\nu = 1$ MHz, $\varepsilon_r^*$, refractive index, $n_D^*$, and density, $\rho^*$, at temperature $T$ and pressure $p = 0.1$ MPa. [a]

| Compound | $\mu$ / D | $T$/K | $\varepsilon_r^*$ Exp. | $\varepsilon_r^*$ Lit. | $n_D^*$ Exp. | $n_D^*$ Lit. | $\rho^*$ / g·cm$^{-3}$ Exp. | $\rho^*$ / g·cm$^{-3}$ Lit. |
|---|---|---|---|---|---|---|---|---|
| methanol | 1.664 [57] | 293.15 | 33.569 | 33.61 [58] | 1.32878 | 1.32859 [59] | 0.79163 | 0.7916 [60] 0.791400 [61] |
| | | 298.15 | 32.619 | 32.62 [58] | 1.32667 | 1.3267 [62] 1.32652 [63] | 0.78695 | 0.7869 [64] 0.786884 [65] |
| | | 303.15 | 31.652 | 31.66 [58] | 1.32457 | 1.32457 [66] 1.32410 [67] | 0.78222 | 0.782158 [65] |
| 1-propanol | 1.629 [57] | 293.15 | 21.146 | 21.15 [68] | 1.38505 | 1.38512 [69] | 0.80366 | 0.80361 [70] |
| | | 298.15 | 20.450 | 20.42 [68] | 1.38304 | 1.38307 [67] | 0.79968 | 0.79960 [70] |
| | | 303.15 | 19.788 | 19.75 [68] | 1.38100 | 1.38104 [67] | 0.79566 | 0.79561 [70] |
| 1-butanol | 1.614 [57] | 293.15 | 18.198 | 18.19 [68] | 1.39925 | 1.3993 [71] | 0.80985 | 0.80982 [72] 0.8098 [73] |
| | | 298.15 | 17.548 | 17.53 [68] | 1.39732 | 1.397336 [74] | 0.80606 | 0.80606 [72] |
| | | 303.15 | 16.927 | 16.89 [68] | 1.39536 | 1.3953 [75] | 0.80222 | 0.8022 [73] |
| 1-pentanol | 1.598 [57] | 293.15 | 15.695 | 15.63 [58] | 1.40992 | 1.40986 [67] | 0.81466 | 0.81468 [76] |
| | | 298.15 | 15.099 | 15.08 [77] | 1.40796 | 1.40789 [67] | 0.81103 | 0.81103 [76] |
| | | 303.15 | 14.523 | 14.44 [58] | 1.40603 | 1.40592 [78] | 0.80735 | 0.81737 [76] |
| 1-heptanol | 1.583 [57] | 293.15 | 12.016 | 11.54 [79] | 1.42422 | 1.42433 [80] | 0.82237 | 0.8223 [81] |
| | | 298.15 | 11.506 | 11.45 [77] | 1.42234 | 1.42240 [80] | 0.81890 | 0.81881 [82] |
| | | 303.15 | 11.021 | 11.07 [47] | 1.42048 | 1.42047 [78] 1.42048 [80] | 0.81537 | 0.8153 [81] |
| di-$n$-propylamine (DPA) | 1.1 [83] | 293.15 | 3.130 | 3.31 [84] 3.068 [45] | 1.40417 | 1.4043 [45] | 0.737782 | 0.7375 [45] |
| | | 298.15 | 3.080 | 3.24 [84] | 1.40154 | 1.40132 [85] | 0.733220 | 0.73321 [50] |
| | | 303.15 | 3.032 | 3.18 [84] | 1.39890 | 1.4022 [86] | 0.728698 | 0.729087 [87] |

[a] The standard uncertainties are: $u(T) = 0.02$ K (for $\rho^*$ measurements, $u(T) = 0.01$ K); $u(p) = 1$ kPa; $u(\nu) = 20$ Hz; $u(n_D^*) = 0.00008$. The relative standard uncertainties are: $u_r(\rho^*) = 0.0012$, $u_r(\varepsilon_r^*) = 0.003$.

Table 3

Volume fractions of 1-alkanol, $\phi_1$, relative permittivities, $\varepsilon_r$, and excess relative permittivities, $\varepsilon_r^E$, of 1-alkanol (1) + di-$n$-propylamine (DPA) (2) mixtures as functions of the mole fraction of the 1-alkanol, $x_1$, at temperature $T$, pressure $p = 0.1$ MPa and frequency $\nu = 1$ MHz. [a]

| $x_1$ | $\phi_1$ | $\varepsilon_r$ | $\varepsilon_r^E$ | $x_1$ | $\phi_1$ | $\varepsilon_r$ | $\varepsilon_r^E$ |
|---|---|---|---|---|---|---|---|
| \multicolumn{8}{c}{methanol (1) + DPA (2) ; $T$/K = 293.15} |
| 0.0000 | 0.0000 | 3.131 |        | 0.5940 | 0.3016 | 14.249 | 1.938 |
| 0.0664 | 0.0206 | 3.698 | −0.060 | 0.6918 | 0.3985 | 17.679 | 2.418 |
| 0.1066 | 0.0340 | 4.093 | −0.073 | 0.7948 | 0.5334 | 21.905 | 2.538 |
| 0.1503 | 0.0496 | 4.582 | −0.059 | 0.8492 | 0.6243 | 24.438 | 2.305 |
| 0.1990 | 0.0683 | 5.201 | −0.009 | 0.9012 | 0.7291 | 27.148 | 1.825 |
| 0.3091 | 0.1166 | 6.956 | 0.276  | 0.9498 | 0.8481 | 30.041 | 1.096 |
| 0.4068 | 0.1683 | 8.965 | 0.711  | 0.9749 | 0.9198 | 31.723 | 0.595 |
| 0.5110 | 0.2357 | 11.698 | 1.393 | 1.0000 | 1.0000 | 33.569 | |
| \multicolumn{8}{c}{methanol (1) + DPA (2) ; $T$/K = 298.15} |
| 0.0000 | 0.0000 | 3.081 |        | 0.5940 | 0.3015 | 13.735 | 1.748 |
| 0.0664 | 0.0206 | 3.620 | −0.069 | 0.6918 | 0.3984 | 17.081 | 2.232 |
| 0.1066 | 0.0340 | 3.998 | −0.087 | 0.7948 | 0.5333 | 21.229 | 2.395 |
| 0.1503 | 0.0496 | 4.466 | −0.080 | 0.8492 | 0.6243 | 23.704 | 2.182 |
| 0.1990 | 0.0683 | 5.054 | −0.044 | 0.9012 | 0.7291 | 26.354 | 1.737 |
| 0.3091 | 0.1166 | 6.725 | 0.200  | 0.9498 | 0.8481 | 29.179 | 1.047 |
| 0.4068 | 0.1683 | 8.641 | 0.589  | 0.9749 | 0.9197 | 30.817 | 0.570 |
| 0.5110 | 0.2357 | 11.261 | 1.218 | 1.0000 | 1.0000 | 32.619 | |
| \multicolumn{8}{c}{methanol (1) + DPA (2) ; $T$/K = 303.15} |
| 0.0000 | 0.0000 | 3.035 |        | 0.5940 | 0.3015 | 13.252 | 1.589 |
| 0.0664 | 0.0205 | 3.552 | −0.070 | 0.6918 | 0.3984 | 16.517 | 2.081 |
| 0.1066 | 0.0340 | 3.913 | −0.095 | 0.7948 | 0.5333 | 20.583 | 2.287 |
| 0.1503 | 0.0496 | 4.357 | −0.097 | 0.8492 | 0.6242 | 22.980 | 2.082 |
| 0.1990 | 0.0683 | 4.920 | −0.070 | 0.9012 | 0.7290 | 25.562 | 1.665 |
| 0.3091 | 0.1166 | 6.514 | 0.142  | 0.9498 | 0.8481 | 28.312 | 1.007 |
| 0.4068 | 0.1683 | 8.346 | 0.495  | 0.9749 | 0.9197 | 29.903 | 0.549 |
| 0.5110 | 0.2356 | 10.861 | 1.084 | 1.0000 | 1.0000 | 31.652 | |
| \multicolumn{8}{c}{1-propanol (1) + DPA (2) ; $T$/K = 293.15} |
| 0.0000 | 0.0000 | 3.130 |        | 0.5948 | 0.4445 | 10.847 | −0.291 |
| 0.0638 | 0.0358 | 3.573 | −0.202 | 0.7018 | 0.5620 | 13.243 | −0.012 |
| 0.0866 | 0.0492 | 3.743 | −0.273 | 0.7967 | 0.6812 | 15.546 | 0.144 |
| 0.1427 | 0.0832 | 4.207 | −0.422 | 0.8431 | 0.7455 | 16.741 | 0.180 |
| 0.2045 | 0.1229 | 4.789 | −0.555 | 0.8993 | 0.8296 | 18.256 | 0.180 |
| 0.2917 | 0.1834 | 5.761 | −0.673 | 0.9487 | 0.9098 | 19.647 | 0.126 |
| 0.3939 | 0.2616 | 7.178 | − 0.665 | 1.0000 | 1.0000 | 21.146 | |
| 0.5012 | 0.3539 | 8.988 | − 0.518 | | | | |

1-propanol (1) + DPA (2) ; $T/K$ = 298.15

| | | | | | | | |
|---|---|---|---|---|---|---|---|
| 0.0000 | 0.0000 | 3.080 | | 0.5948 | 0.4442 | 10.419 | − 0.377 |
| 0.0638 | 0.0358 | 3.501 | − 0.201 | 0.7018 | 0.5617 | 12.739 | − 0.098 |
| 0.0866 | 0.0491 | 3.665 | − 0.268 | 0.7967 | 0.6809 | 14.969 | 0.062 |
| 0.1427 | 0.0831 | 4.108 | − 0.415 | 0.8431 | 0.7453 | 16.134 | 0.108 |
| 0.2045 | 0.1228 | 4.660 | − 0.553 | 0.8993 | 0.8294 | 17.623 | 0.136 |
| 0.2917 | 0.1832 | 5.580 | − 0.682 | 0.9487 | 0.9097 | 18.987 | 0.106 |
| 0.3939 | 0.2614 | 6.929 | − 0.692 | 1.0000 | 1.0000 | 20.450 | |
| 0.5012 | 0.3536 | 8.647 | − 0.575 | | | | |

1-propanol (1) + DPA (2) ; $T/K$ = 303.15

| | | | | | | | |
|---|---|---|---|---|---|---|---|
| 0.0000 | 0.0000 | 3.032 | | 0.5948 | 0.4440 | 10.029 | − 0.443 |
| 0.0638 | 0.0357 | 3.439 | − 0.191 | 0.7018 | 0.5614 | 12.260 | − 0.179 |
| 0.0866 | 0.0490 | 3.593 | − 0.260 | 0.7967 | 0.6807 | 14.433 | − 0.005 |
| 0.1427 | 0.0830 | 4.016 | − 0.407 | 0.8431 | 0.7451 | 15.567 | 0.050 |
| 0.2045 | 0.1227 | 4.540 | − 0.548 | 0.8993 | 0.8293 | 17.021 | 0.093 |
| 0.2917 | 0.1830 | 5.417 | − 0.681 | 0.9487 | 0.9096 | 18.361 | 0.088 |
| 0.3939 | 0.2612 | 6.698 | − 0.711 | 1.0000 | 1.0000 | 19.788 | |
| 0.5012 | 0.3534 | 8.331 | − 0.623 | | | | |

1-butanol (1) + DPA (2) ; $T/K$ = 293.15

| | | | | | | | |
|---|---|---|---|---|---|---|---|
| 0.0000 | 0.0000 | 3.132 | | 0.5991 | 0.4993 | 9.972 | − 0.682 |
| 0.0484 | 0.0328 | 3.455 | − 0.171 | 0.6896 | 0.5972 | 11.635 | − 0.494 |
| 0.1063 | 0.0735 | 3.872 | − 0.367 | 0.7484 | 0.6650 | 12.765 | − 0.386 |
| 0.1418 | 0.0993 | 4.146 | − 0.482 | 0.8041 | 0.7326 | 13.891 | − 0.278 |
| 0.2157 | 0.1551 | 4.788 | − 0.681 | 0.8418 | 0.7803 | 14.686 | − 0.202 |
| 0.3006 | 0.2229 | 5.660 | − 0.830 | 0.8890 | 0.8424 | 15.681 | − 0.143 |
| 0.4064 | 0.3136 | 6.961 | − 0.896 | 0.9514 | 0.9289 | 17.070 | − 0.057 |
| 0.5030 | 0.4031 | 8.373 | − 0.832 | 1.0000 | 1.0000 | 18.198 | |

1-butanol (1) + DPA (2) ; $T/K$ = 298.15

| | | | | | | | |
|---|---|---|---|---|---|---|---|
| 0.0000 | 0.0000 | 3.082 | | 0.5991 | 0.4989 | 9.580 | − 0.719 |
| 0.0484 | 0.0328 | 3.391 | − 0.165 | 0.6896 | 0.5968 | 11.179 | − 0.536 |
| 0.1063 | 0.0734 | 3.786 | − 0.358 | 0.7484 | 0.6646 | 12.268 | − 0.428 |
| 0.1418 | 0.0992 | 4.049 | − 0.468 | 0.8041 | 0.7323 | 13.368 | − 0.307 |
| 0.2157 | 0.1549 | 4.659 | − 0.664 | 0.8418 | 0.7800 | 14.125 | − 0.240 |
| 0.3006 | 0.2226 | 5.486 | − 0.816 | 0.8890 | 0.8422 | 15.113 | − 0.152 |
| 0.4064 | 0.3133 | 6.718 | − 0.896 | 0.9514 | 0.9288 | 16.455 | − 0.063 |
| 0.5030 | 0.4028 | 8.057 | − 0.852 | 1.0000 | 1.0000 | 17.548 | |

1-butanol (1) + DPA (2) ; $T/K$ = 303.15

| | | | | | | | |
|---|---|---|---|---|---|---|---|
| 0.0000 | 0.0000 | 3.036 | | 0.5991 | 0.4986 | 9.223 | − 0.739 |
| 0.0484 | 0.0327 | 3.330 | − 0.160 | 0.6896 | 0.5965 | 10.738 | − 0.584 |
| 0.1063 | 0.0733 | 3.710 | − 0.344 | 0.7484 | 0.6643 | 11.811 | − 0.453 |
| 0.1418 | 0.0990 | 3.961 | − 0.450 | 0.8041 | 0.7320 | 12.869 | − 0.335 |
| 0.2157 | 0.1547 | 4.543 | − 0.642 | 0.8418 | 0.7798 | 13.609 | − 0.259 |
| 0.3006 | 0.2224 | 5.329 | − 0.796 | 0.8890 | 0.8420 | 14.568 | − 0.164 |
| 0.4064 | 0.3130 | 6.497 | − 0.887 | 0.9514 | 0.9287 | 15.872 | − 0.065 |
| 0.5030 | 0.4024 | 7.770 | − 0.856 | 1.0000 | 1.0000 | 16.927 | |
| | | 1-pentanol (1) + DPA (2) ; $T$/K = 293.15 | | | | | |
| 0.0000 | 0.0000 | 3.130 | | 0.5985 | 0.5404 | 9.078 | − 0.842 |
| 0.0530 | 0.0423 | 3.469 | − 0.192 | 0.6570 | 0.6018 | 9.926 | − 0.766 |
| 0.1086 | 0.0877 | 3.850 | − 0.382 | 0.6985 | 0.6464 | 10.555 | − 0.697 |
| 0.1497 | 0.1220 | 4.157 | − 0.506 | 0.7450 | 0.6974 | 11.282 | − 0.611 |
| 0.2032 | 0.1675 | 4.581 | − 0.654 | 0.7921 | 0.7504 | 12.039 | − 0.520 |
| 0.2597 | 0.2168 | 5.076 | − 0.778 | 0.8447 | 0.8110 | 12.918 | − 0.402 |
| 0.3006 | 0.2532 | 5.461 | − 0.850 | 0.9027 | 0.8798 | 13.918 | − 0.267 |
| 0.4064 | 0.3507 | 6.590 | − 0.947 | 0.9435 | 0.9294 | 14.657 | − 0.151 |
| 0.4882 | 0.4294 | 7.589 | − 0.936 | 1.0000 | 1.0000 | 15.695 | |
| 0.5421 | 0.4829 | 8.290 | − 0.908 | | | | |
| | | 1-pentanol (1) + DPA (2) ; $T$/K = 298.15 | | | | | |
| 0.0000 | 0.0000 | 3.077 | | 0.5985 | 0.5400 | 8.729 | − 0.840 |
| 0.0530 | 0.0422 | 3.402 | − 0.182 | 0.6570 | 0.6014 | 9.540 | − 0.767 |
| 0.1086 | 0.0875 | 3.767 | − 0.362 | 0.6985 | 0.6460 | 10.144 | − 0.699 |
| 0.1497 | 0.1218 | 4.062 | − 0.479 | 0.7450 | 0.6970 | 10.849 | − 0.607 |
| 0.2032 | 0.1672 | 4.465 | − 0.622 | 0.7921 | 0.7500 | 11.574 | − 0.519 |
| 0.2597 | 0.2165 | 4.935 | − 0.745 | 0.8447 | 0.8107 | 12.421 | − 0.402 |
| 0.3006 | 0.2529 | 5.303 | − 0.814 | 0.9027 | 0.8796 | 13.397 | − 0.255 |
| 0.4064 | 0.3503 | 6.367 | − 0.921 | 0.9435 | 0.9293 | 14.100 | − 0.149 |
| 0.4882 | 0.4290 | 7.316 | − 0.918 | 1.0000 | 1.0000 | 15.099 | |
| 0.5421 | 0.4825 | 7.978 | − 0.900 | | | | |
| | | 1-pentanol (1) + DPA (2) ; $T$/K = 303.15 | | | | | |
| 0.0000 | 0.0000 | 3.031 | | 0.5985 | 0.5396 | 8.413 | − 0.819 |
| 0.0530 | 0.0421 | 3.342 | − 0.173 | 0.6570 | 0.6010 | 9.187 | − 0.751 |
| 0.1086 | 0.0874 | 3.692 | − 0.343 | 0.6985 | 0.6456 | 9.765 | − 0.685 |
| 0.1497 | 0.1216 | 3.972 | − 0.456 | 0.7450 | 0.6967 | 10.436 | − 0.601 |
| 0.2032 | 0.1670 | 4.357 | − 0.593 | 0.7921 | 0.7497 | 11.139 | − 0.508 |
| 0.2597 | 0.2162 | 4.805 | − 0.711 | 0.8447 | 0.8105 | 11.962 | − 0.383 |
| 0.3006 | 0.2526 | 5.156 | − 0.778 | 0.9027 | 0.8794 | 12.894 | − 0.243 |
| 0.4064 | 0.3499 | 6.167 | − 0.885 | 0.9435 | 0.9292 | 13.570 | − 0.139 |
| 0.4882 | 0.4286 | 7.063 | − 0.893 | 1.0000 | 1.0000 | 14.523 | |
| 0.5421 | 0.4821 | 7.695 | − 0.876 | | | | |
| | | 1-heptanol (1) + DPA (2) ; $T$/K = 293.15 | | | | | |
| 0.0000 | 0.0000 | 3.131 | | 0.5883 | 0.5955 | 7.677 | − 0.745 |
| 0.0529 | 0.0544 | 3.454 | − 0.160 | 0.6378 | 0.6446 | 8.154 | − 0.704 |
| 0.1003 | 0.1030 | 3.745 | − 0.301 | 0.6965 | 0.7028 | 8.735 | − 0.640 |
| 0.1474 | 0.1512 | 4.050 | − 0.424 | 0.7424 | 0.7481 | 9.197 | − 0.581 |
| 0.1990 | 0.2038 | 4.401 | − 0.541 | 0.7892 | 0.7941 | 9.687 | − 0.500 |
| 0.2471 | 0.2527 | 4.748 | − 0.628 | 0.8426 | 0.8465 | 10.250 | − 0.402 |
| 0.2924 | 0.2986 | 5.083 | − 0.701 | 0.8913 | 0.8941 | 10.777 | − 0.298 |
| 0.3427 | 0.3494 | 5.485 | − 0.750 | 0.9395 | 0.9412 | 11.325 | − 0.169 |
| 0.3943 | 0.4014 | 5.915 | − 0.782 | 1.0000 | 1.0000 | 12.016 | |
| 0.4929 | 0.5003 | 6.783 | − 0.793 | | | | |
| 1-heptanol (1) + DPA (2) ; $T$/K = 298.15 | | | | | | | |
| 0.0000 | 0.0000 | 3.080 | | 0.5883 | 0.5950 | 7.390 | − 0.703 |
| 0.0529 | 0.0543 | 3.389 | − 0.149 | 0.6378 | 0.6442 | 7.847 | − 0.661 |
| 0.1003 | 0.1028 | 3.669 | − 0.277 | 0.6965 | 0.7023 | 8.397 | − 0.601 |
| 0.1474 | 0.1509 | 3.957 | − 0.394 | 0.7424 | 0.7477 | 8.837 | − 0.543 |
| 0.1990 | 0.2035 | 4.295 | − 0.500 | 0.7892 | 0.7938 | 9.307 | − 0.462 |
| 0.2471 | 0.2523 | 4.626 | − 0.580 | 0.8426 | 0.8463 | 9.838 | − 0.373 |
| 0.2924 | 0.2982 | 4.941 | − 0.652 | 0.8913 | 0.8940 | 10.344 | − 0.269 |
| 0.3427 | 0.3490 | 5.325 | − 0.696 | 0.9395 | 0.9411 | 10.854 | − 0.156 |
| 0.3943 | 0.4010 | 5.728 | − 0.731 | 1.0000 | 1.0000 | 11.506 | |
| 0.4929 | 0.4998 | 6.547 | − 0.744 | | | | |
| 1-heptanol (1) + DPA (2) ; $T$/K = 303.15 | | | | | | | |
| 0.0000 | 0.0000 | 3.033 | | 0.5883 | 0.5946 | 7.128 | − 0.655 |
| 0.0529 | 0.0542 | 3.329 | − 0.137 | 0.6378 | 0.6438 | 7.565 | − 0.611 |
| 0.1003 | 0.1027 | 3.597 | − 0.256 | 0.6965 | 0.7020 | 8.086 | − 0.555 |
| 0.1474 | 0.1507 | 3.876 | − 0.361 | 0.7424 | 0.7473 | 8.497 | − 0.505 |
| 0.1990 | 0.2032 | 4.196 | − 0.460 | 0.7892 | 0.7935 | 8.956 | − 0.415 |
| 0.2471 | 0.2520 | 4.512 | − 0.534 | 0.8426 | 0.8460 | 9.459 | − 0.332 |
| 0.2924 | 0.2978 | 4.812 | − 0.600 | 0.8913 | 0.8938 | 9.923 | − 0.250 |
| 0.3427 | 0.3486 | 5.177 | − 0.641 | 0.9395 | 0.9410 | 10.415 | − 0.135 |
| 0.3943 | 0.4005 | 5.558 | − 0.674 | 1.0000 | 1.0000 | 11.021 | |
| 0.4929 | 0.4994 | 6.334 | − 0.688 | | | | |

[a]The standard uncertainties are: $u(T)=0.02$ K; $u(p)=1$ kPa; $u(\nu)=20$ Hz; $u(x_1)=0.0010$; $u(\phi_1)=0.004$. The relative standard uncertainty is: $u_r(\varepsilon_r)=0.003$; and the relative combined expanded uncertainty (0.95 level of confidence) is $U_{rc}(\varepsilon_r^E)=0.03$.

Table 4

Volume fractions of 1-alkanol, $\phi_1$, refractive indices, $n_D$, and excess refractive indices, $n_D^E$, of 1-alkanol (1) + di-$n$-propylamine (DPA) (2) mixtures as functions of the mole fraction of the 1-alkanol, $x_1$, at temperature $T$ and pressure $p = 0.1$ MPa. [a]

| $x_1$ | $\phi_1$ | $n_D$ | $10^5 n_D^E$ | $x_1$ | $\phi_1$ | $n_D$ | $10^5 n_D^E$ |
|---|---|---|---|---|---|---|---|
| \multicolumn{8}{c}{methanol (1) + DPA (2) ; $T$/K = 293.15} |
| 0.0000 | 0.0000 | 1.40417 |  | 0.5900 | 0.2981 | 1.39077 | 884 |
| 0.0374 | 0.0113 | 1.40406 | 72 | 0.6894 | 0.3958 | 1.38332 | 880 |
| 0.1111 | 0.0356 | 1.40355 | 199 | 0.8067 | 0.5519 | 1.36969 | 715 |
| 0.1464 | 0.0482 | 1.40324 | 261 | 0.8492 | 0.6243 | 1.36312 | 611 |
| 0.2203 | 0.0770 | 1.40238 | 387 | 0.8989 | 0.7241 | 1.35397 | 446 |
| 0.3034 | 0.1139 | 1.40103 | 524 | 0.9498 | 0.8481 | 1.34258 | 234 |
| 0.4174 | 0.1745 | 1.39825 | 694 | 0.9829 | 0.9443 | 1.33382 | 82 |
| 0.4937 | 0.2235 | 1.39553 | 785 | 1.0000 | 1.0000 | 1.32878 |  |
| \multicolumn{8}{c}{methanol (1) + DPA (2) ; $T$/K = 298.15} |
| 0.0000 | 0.0000 | 1.40154 |  | 0.5900 | 0.2980 | 1.38830 | 864 |
| 0.0374 | 0.0113 | 1.40142 | 70 | 0.6894 | 0.3957 | 1.38108 | 849 |
| 0.1111 | 0.0356 | 1.40093 | 199 | 0.8067 | 0.5518 | 1.36764 | 661 |
| 0.1464 | 0.0482 | 1.40056 | 254 | 0.8492 | 0.6243 | 1.36109 | 552 |
| 0.2203 | 0.0769 | 1.39974 | 381 | 0.8989 | 0.7240 | 1.35202 | 397 |
| 0.3034 | 0.1139 | 1.39844 | 522 | 0.9498 | 0.8481 | 1.34058 | 208 |
| 0.4174 | 0.1745 | 1.39573 | 696 | 0.9829 | 0.9443 | 1.33174 | 73 |
| 0.4937 | 0.2234 | 1.39316 | 799 | 1.0000 | 1.0000 | 1.32667 |  |
| \multicolumn{8}{c}{methanol (1) + DPA (2) ; $T$/K = 303.15} |
| 0.0000 | 0.0000 | 1.39890 |  | 0.5900 | 0.2980 | 1.38601 | 865 |
| 0.0374 | 0.0113 | 1.39881 | 73 | 0.6894 | 0.3957 | 1.37877 | 868 |
| 0.1111 | 0.0356 | 1.39834 | 202 | 0.8067 | 0.5518 | 1.36554 | 690 |
| 0.1464 | 0.0482 | 1.39800 | 259 | 0.8492 | 0.6242 | 1.35909 | 581 |
| 0.2203 | 0.0769 | 1.39719 | 387 | 0.8989 | 0.7240 | 1.34996 | 427 |
| 0.3034 | 0.1139 | 1.39590 | 527 | 0.9498 | 0.8481 | 1.33847 | 227 |
| 0.4174 | 0.1745 | 1.39317 | 695 | 0.9829 | 0.9443 | 1.32964 | 79 |
| 0.4937 | 0.2234 | 1.39071 | 807 | 1.0000 | 1.0000 | 1.32457 |  |
| \multicolumn{8}{c}{1-propanol (1) + DPA (2) ; $T$/K = 293.15} |
| 0.0000 | 0.0000 | 1.40417 |  | 0.6004 | 0.4503 | 1.40172 | 613 |
| 0.0465 | 0.0259 | 1.40431 | 63 | 0.6962 | 0.5554 | 1.39932 | 574 |
| 0.0897 | 0.0510 | 1.40448 | 128 | 0.7952 | 0.6792 | 1.39596 | 475 |
| 0.1524 | 0.0893 | 1.40466 | 219 | 0.8476 | 0.7520 | 1.39367 | 385 |
| 0.2024 | 0.1215 | 1.40471 | 285 | 0.8965 | 0.8252 | 1.39125 | 284 |
| 0.3121 | 0.1983 | 1.40456 | 416 | 0.9575 | 0.9247 | 1.38780 | 130 |
| 0.4059 | 0.2714 | 1.40413 | 512 | 1.0000 | 1.0000 | 1.38505 | |
| 0.4936 | 0.3470 | 1.40333 | 577 | | | | |

1-propanol (1) + DPA (2) ; $T/K = 298.15$

| | | | | | | | |
|---|---|---|---|---|---|---|---|
| 0.0000 | 0.0000 | 1.40159 | | 0.6004 | 0.4500 | 1.39924 | 597 |
| 0.0465 | 0.0259 | 1.40173 | 62 | 0.6962 | 0.5551 | 1.39697 | 565 |
| 0.0897 | 0.0509 | 1.40191 | 126 | 0.7952 | 0.6789 | 1.39376 | 474 |
| 0.1524 | 0.0892 | 1.40207 | 212 | 0.8476 | 0.7518 | 1.39153 | 386 |
| 0.2024 | 0.1214 | 1.40220 | 285 | 0.8965 | 0.8251 | 1.38914 | 284 |
| 0.3121 | 0.1981 | 1.40205 | 412 | 0.9575 | 0.9246 | 1.38576 | 131 |
| 0.4059 | 0.2712 | 1.40163 | 505 | 1.0000 | 1.0000 | 1.38304 | |
| 0.4936 | 0.3467 | 1.40092 | 573 | | | | |

1-propanol (1) + DPA (2) ; $T/K = 303.15$

| | | | | | | | |
|---|---|---|---|---|---|---|---|
| 0.0000 | 0.0000 | 1.39890 | | 0.6004 | 0.4497 | 1.39684 | 596 |
| 0.0465 | 0.0258 | 1.39904 | 60 | 0.6962 | 0.5548 | 1.39460 | 560 |
| 0.0897 | 0.0509 | 1.39921 | 122 | 0.7952 | 0.6786 | 1.39143 | 465 |
| 0.1524 | 0.0891 | 1.39941 | 210 | 0.8476 | 0.7516 | 1.38929 | 382 |
| 0.2024 | 0.1213 | 1.39951 | 277 | 0.8965 | 0.8249 | 1.38698 | 283 |
| 0.3121 | 0.1979 | 1.39942 | 404 | 0.9575 | 0.9245 | 1.38362 | 126 |
| 0.4059 | 0.2709 | 1.39904 | 497 | 1.0000 | 1.0000 | 1.38100 | |
| 0.4936 | 0.3465 | 1.39835 | 563 | | | | |

1-butanol (1) + DPA (2) ; $T/K = 293.15$

| | | | | | | | |
|---|---|---|---|---|---|---|---|
| 0.0000 | 0.0000 | 1.40409 | | 0.5978 | 0.4980 | 1.40746 | 578 |
| 0.0519 | 0.0352 | 1.40477 | 85 | 0.6903 | 0.5980 | 1.40661 | 541 |
| 0.1199 | 0.0833 | 1.40556 | 187 | 0.7948 | 0.7210 | 1.40492 | 432 |
| 0.1581 | 0.1114 | 1.40591 | 236 | 0.8518 | 0.7932 | 1.40366 | 341 |
| 0.2220 | 0.1600 | 1.40653 | 321 | 0.8875 | 0.8404 | 1.40274 | 272 |
| 0.3097 | 0.2304 | 1.40713 | 415 | 0.9433 | 0.9174 | 1.40111 | 146 |
| 0.3936 | 0.3022 | 1.40759 | 496 | 1.0000 | 1.0000 | 1.39925 | |
| 0.5007 | 0.4009 | 1.40773 | 558 | | | | |

1-butanol (1) + DPA (2) ; $T/K = 298.15$

| | | | | | | | |
|---|---|---|---|---|---|---|---|
| 0.0000 | 0.0000 | 1.40147 | | 0.5978 | 0.4976 | 1.40506 | 565 |
| 0.0519 | 0.0352 | 1.40216 | 84 | 0.6903 | 0.5976 | 1.40427 | 528 |
| 0.1199 | 0.0832 | 1.40293 | 180 | 0.7948 | 0.7207 | 1.40268 | 420 |
| 0.1581 | 0.1112 | 1.40325 | 224 | 0.8518 | 0.7929 | 1.40148 | 330 |
| 0.2220 | 0.1598 | 1.40396 | 315 | 0.8875 | 0.8402 | 1.40059 | 261 |
| 0.3097 | 0.2301 | 1.40458 | 406 | 0.9433 | 0.9173 | 1.39903 | 137 |
| 0.3936 | 0.3019 | 1.40504 | 482 | 1.0000 | 1.0000 | 1.39732 | |
| 0.5007 | 0.4005 | 1.40529 | 548 | | | | |

1-butanol (1) + DPA (2) ; $T/K = 303.15$

| | | | | | | | |
|---|---|---|---|---|---|---|---|
| 0.0000 | 0.0000 | 1.39888 | | 0.5978 | 0.4972 | 1.40264 | 551 |
| 0.0519 | 0.0351 | 1.39950 | 74 | 0.6903 | 0.5973 | 1.40199 | 521 |
| 0.1199 | 0.0831 | 1.40031 | 172 | 0.7948 | 0.7204 | 1.40048 | 413 |
| 0.1581 | 0.1111 | 1.40067 | 218 | 0.8518 | 0.7927 | 1.39930 | 321 |
| 0.2220 | 0.1596 | 1.40138 | 306 | 0.8875 | 0.8400 | 1.39848 | 256 |
| 0.3097 | 0.2299 | 1.40206 | 399 | 0.9433 | 0.9171 | 1.39698 | 133 |
| 0.3936 | 0.3016 | 1.40253 | 471 | 1.0000 | 1.0000 | 1.39536 | |
| 0.5007 | 0.4002 | 1.40284 | 537 | | | | |
| | | 1-pentanol (1) + DPA (2) ; $T/K = 293.15$ | | | | | |
| 0.0000 | 0.0000 | 1.40409 | | 0.6080 | 0.5503 | 1.41265 | 535 |
| 0.0523 | 0.0417 | 1.40515 | 82 | 0.7040 | 0.6523 | 1.41274 | 484 |
| 0.0950 | 0.0765 | 1.40602 | 148 | 0.7909 | 0.7490 | 1.41238 | 392 |
| 0.1538 | 0.1254 | 1.40714 | 232 | 0.8562 | 0.8245 | 1.41188 | 298 |
| 0.2103 | 0.1736 | 1.40815 | 305 | 0.8980 | 0.8741 | 1.41142 | 223 |
| 0.2984 | 0.2512 | 1.40965 | 409 | 0.9497 | 0.9371 | 1.41074 | 119 |
| 0.4033 | 0.3478 | 1.41103 | 491 | 1.0000 | 1.0000 | 1.40992 | |
| 0.5068 | 0.4477 | 1.41207 | 537 | | | | |
| | | 1-pentanol (1) + DPA (2) ; $T/K = 298.15$ | | | | | |
| 0.0000 | 0.0000 | 1.40147 | | 0.6080 | 0.5498 | 1.41045 | 541 |
| 0.0523 | 0.0417 | 1.40255 | 81 | 0.7040 | 0.6519 | 1.41065 | 495 |
| 0.0950 | 0.0764 | 1.40342 | 145 | 0.7909 | 0.7487 | 1.41035 | 402 |
| 0.1538 | 0.1252 | 1.40462 | 234 | 0.8562 | 0.8242 | 1.40987 | 305 |
| 0.2103 | 0.1734 | 1.40562 | 302 | 0.8980 | 0.8740 | 1.40943 | 229 |
| 0.2984 | 0.2509 | 1.40715 | 405 | 0.9497 | 0.9370 | 1.40875 | 120 |
| 0.4033 | 0.3474 | 1.40861 | 488 | 1.0000 | 1.0000 | 1.40796 | |
| 0.5068 | 0.4473 | 1.40974 | 536 | | | | |
| | | 1-pentanol (1) + DPA (2) ; $T/K = 303.15$ | | | | | |
| 0.0000 | 0.0000 | 1.39888 | | 0.6080 | 0.5494 | 1.40811 | 530 |
| 0.0523 | 0.0416 | 1.39999 | 81 | 0.7040 | 0.6516 | 1.40838 | 484 |
| 0.0950 | 0.0762 | 1.40087 | 144 | 0.7909 | 0.7484 | 1.40813 | 390 |
| 0.1538 | 0.1250 | 1.40208 | 230 | 0.8562 | 0.8240 | 1.40777 | 300 |
| 0.2103 | 0.1731 | 1.40316 | 304 | 0.8980 | 0.8738 | 1.40728 | 215 |
| 0.2984 | 0.2506 | 1.40468 | 400 | 0.9497 | 0.9369 | 1.40671 | 113 |
| 0.4033 | 0.3470 | 1.40617 | 480 | 1.0000 | 1.0000 | 1.40603 | |
| 0.5068 | 0.4469 | 1.40739 | 531 | | | | |
| | | 1-heptanol (1) + DPA (2) ; $T/K = 293.15$ | | | | | |
| 0.0000 | 0.0000 | 1.40409 | | 0.5988 | 0.6079 | 1.42105 | 469 |
| 0.0533 | 0.0553 | 1.40617 | 96 | 0.6970 | 0.7050 | 1.42249 | 418 |
| 0.1009 | 0.1044 | 1.40794 | 173 | 0.7957 | 0.8018 | 1.42346 | 321 |
| 0.1434 | 0.1482 | 1.40944 | 235 | 0.8458 | 0.8507 | 1.42381 | 258 |
| 0.1983 | 0.2044 | 1.41129 | 306 | 0.8956 | 0.8991 | 1.42409 | 189 |
| 0.2939 | 0.3019 | 1.41418 | 398 | 0.9427 | 0.9447 | 1.42420 | 109 |
| 0.3915 | 0.4006 | 1.41677 | 458 | 1.0000 | 1.0000 | 1.42422 | |
| 0.4925 | 0.5021 | 1.41906 | 483 | | | | |
| | | 1-heptanol (1) + DPA (2) ; $T/K = 298.15$ | | | | | |
| 0.0000 | 0.0000 | 1.40147 | | 0.5988 | 0.6055 | 1.41883 | 469 |
| 0.0533 | 0.0547 | 1.40355 | 93 | 0.6970 | 0.7028 | 1.42035 | 422 |
| 0.1009 | 0.1034 | 1.40532 | 168 | 0.7957 | 0.8002 | 1.42142 | 323 |
| 0.1434 | 0.1468 | 1.40684 | 229 | 0.8458 | 0.8494 | 1.42180 | 258 |
| 0.1983 | 0.2028 | 1.40873 | 300 | 0.8956 | 0.8982 | 1.42209 | 186 |
| 0.2939 | 0.2997 | 1.41170 | 394 | 0.9427 | 0.9442 | 1.42226 | 108 |
| 0.3915 | 0.3981 | 1.41433 | 451 | 1.0000 | 1.0000 | 1.42234 | |
| 0.4925 | 0.4994 | 1.41673 | 480 | | | | |
| | | 1-heptanol (1) + DPA (2) ; $T/K = 303.15$ | | | | | |
| 0.0000 | 0.0000 | 1.39888 | | 0.5988 | 0.6030 | 1.41664 | 470 |
| 0.0533 | 0.0542 | 1.40101 | 95 | 0.6970 | 0.7007 | 1.41822 | 417 |
| 0.1009 | 0.1025 | 1.40283 | 172 | 0.7957 | 0.7985 | 1.41935 | 320 |
| 0.1434 | 0.1456 | 1.40436 | 231 | 0.8458 | 0.8481 | 1.41980 | 258 |
| 0.1983 | 0.2011 | 1.40628 | 303 | 0.8956 | 0.8972 | 1.42012 | 185 |
| 0.2939 | 0.2975 | 1.40933 | 399 | 0.9427 | 0.9436 | 1.42034 | 107 |
| 0.3915 | 0.3957 | 1.41205 | 458 | 1.0000 | 1.0000 | 1.42048 | |
| 0.4925 | 0.4969 | 1.41451 | 486 | | | | |

[a]The standard uncertainties are: $u(T) = 0.02$ K; $u(p) = 1$ kPa; $u(x_1) = 0.0010$; $u(\phi_1) = 0.004$, $u(n_D) = 0.00008$. The combined expanded uncertainty (0.95 level of confidence) is $U_{rc}(n_D^E) = 0.0002$.

Table 5

Coefficients $A_i$ and standard deviations, $\sigma(F^E)$ (equation (6)), for the representation of $F^E$ at temperature $T$ and pressure $p = 0.1$ MPa for 1-alkanol (1) + di-$n$-propylamine (DPA) (2) systems by equation (5).

| Property $F^E$ | 1-alkanol | $T$/K | $A_0$ | $A_1$ | $A_2$ | $A_3$ | $\sigma(F^E)$ |
|---|---|---|---|---|---|---|---|
| $\varepsilon_r^E$ | methanol | 293.15 | 5.25 | 13.23 | 7.18 | | 0.010 |
| | | 298.15 | 4.60 | 12.65 | 7.4 | | 0.012 |
| | | 303.15 | 4.08 | 12.16 | 7.6 | | 0.016 |
| | 1-propanol | 293.15 | −2.03 | 4.03 | 2.12 | −1.0 | 0.008 |
| | | 298.15 | −2.27 | 3.48 | 2.12 | −0.7 | 0.010 |
| | | 303.15 | −2.46 | 2.81 | 2.1 | | 0.011 |
| | 1-butanol | 293.15 | −3.32 | 2.29 | 1.15 | −1.1 | 0.006 |
| | | 298.15 | −3.40 | 1.89 | 1.23 | −0.78 | 0.005 |
| | | 303.15 | −3.433 | 1.49 | 1.29 | −0.39 | 0.003 |
| | 1-pentanol | 293.15 | −3.738 | 1.02 | 0.43 | −0.71 | 0.004 |
| | | 298.15 | −3.680 | 0.77 | 0.56 | −0.45 | 0.004 |
| | | 303.15 | −3.574 | 0.59 | 0.64 | −0.20 | 0.003 |
| | 1-heptanol | 293.15 | −3.178 | 0.51 | −0.05 | −0.53 | 0.004 |
| | | 298.15 | −2.982 | 0.41 | 0.06 | −0.36 | 0.004 |
| | | 303.15 | −2.76 | 0.35 | 0.14 | −0.2 | 0.006 |
| $10^5 n_D^E$ | methanol | 293.15 | 3200 | 2053 | 24 | −945 | 5 |
| | | 298.15 | 3218 | 2133 | 191 | −712 | 4 |
| | | 303.15 | 3247 | 2236 | 359 | −712 | 4 |
| | 1-propanol | 293.15 | 2328 | 970 | | | 4 |
| | | 298.15 | 2301 | 970 | | | 4 |
| | | 303.15 | 2269 | 977 | | | 4 |
| | 1-butanol | 293.15 | 2240 | 810 | −1 | −316 | 1.7 |
| | | 298.15 | 2195 | 803 | −65 | −356 | 3 |
| | | 303.15 | 2154 | 782 | −104 | −293 | 4 |
| | 1-pentanol | 293.15 | 2126 | 459 | | | 4 |
| | | 298.15 | 2138 | 520 | | | 4 |
| | | 303.15 | 2102 | 470 | | | 5 |
| | 1-heptanol | 293.15 | 1944 | 59 | | | 1.9 |
| | | 298.15 | 1931 | 96 | | | 3 |
| | | 303.15 | 1942 | 61 | | | 1.3 |
| $\left(\dfrac{\partial \varepsilon_r^E}{\partial T}\right)_p$ / K$^{-1}$ | methanol | 298.15 | −0.117 | −0.13 | 0.04 | 0.05 | 0.0011 |
| | 1-propanol | 298.15 | −0.043 | −0.104 | 0.002 | 0.056 | 0.0004 |
| | 1-butanol | 298.15 | −0.011 | −0.080 | 0.013 | 0.08 | 0.0007 |
| | 1-pentanol | 298.15 | 0.0164 | −0.043 | 0.021 | 0.051 | 0.0002 |
| | 1-heptanol | 298.15 | 0.0415 | −0.016 | 0.019 | 0.032 | 0.0004 |

Table 6

Values of the derivative of permittivity with respect to temperature at 298.15 K for pure compounds[a], $\left(\partial \varepsilon_r^* / \partial T\right)_p$, and for mixtures, $\left(\partial \varepsilon_r / \partial T\right)_p$, at $\phi_1 = 0.5$.

| Compound | $\left(\partial \varepsilon_r^* / \partial T\right)_p$ /K$^{-1}$ | | $\left(\partial \varepsilon_r / \partial T\right)_p$ /K$^{-1}$ | |
| --- | --- | --- | --- | --- |
| | Exp. | Lit. | 1-alkanol + DPA | 1-alkanol + HxA |
| Methanol | − 0.192 | −0.195 [68] | − 0.131 | − 0.110 |
| 1-propanol | − 0.136 | −0.130 [88] | − 0.094 | − 0.076 |
| 1-butanol | − 0.127 | −0.122 [88] | − 0.077 | − 0.060 |
| 1-pentanol | − 0.117 | −0.110 [88] | − 0.062 | − 0.044 |
| 1-heptanol | − 0.099 | −0.096 [88] | − 0.044 | − 0.023 |
| DPA | − 0.012 | | | |
| HXA | − 0.0098 | | | |

[a] *n*-hexylamine (HxA), di-*n*-propylamine (DPA).

**Figure 1**: Excess relative permittivities, $\varepsilon_r^E$, of 1-alkanol (1) + di-*n*-propylamine (2) systems at 0.1 MPa, 298.15 K and 1 MHz. Full symbols, experimental values (this work): (●), methanol; (■), 1-propanol; (▲), 1-butanol; (♦), 1-pentanol; (▼), 1-heptanol. Solid lines, calculations with equation (5) using the coefficients from Table 5.

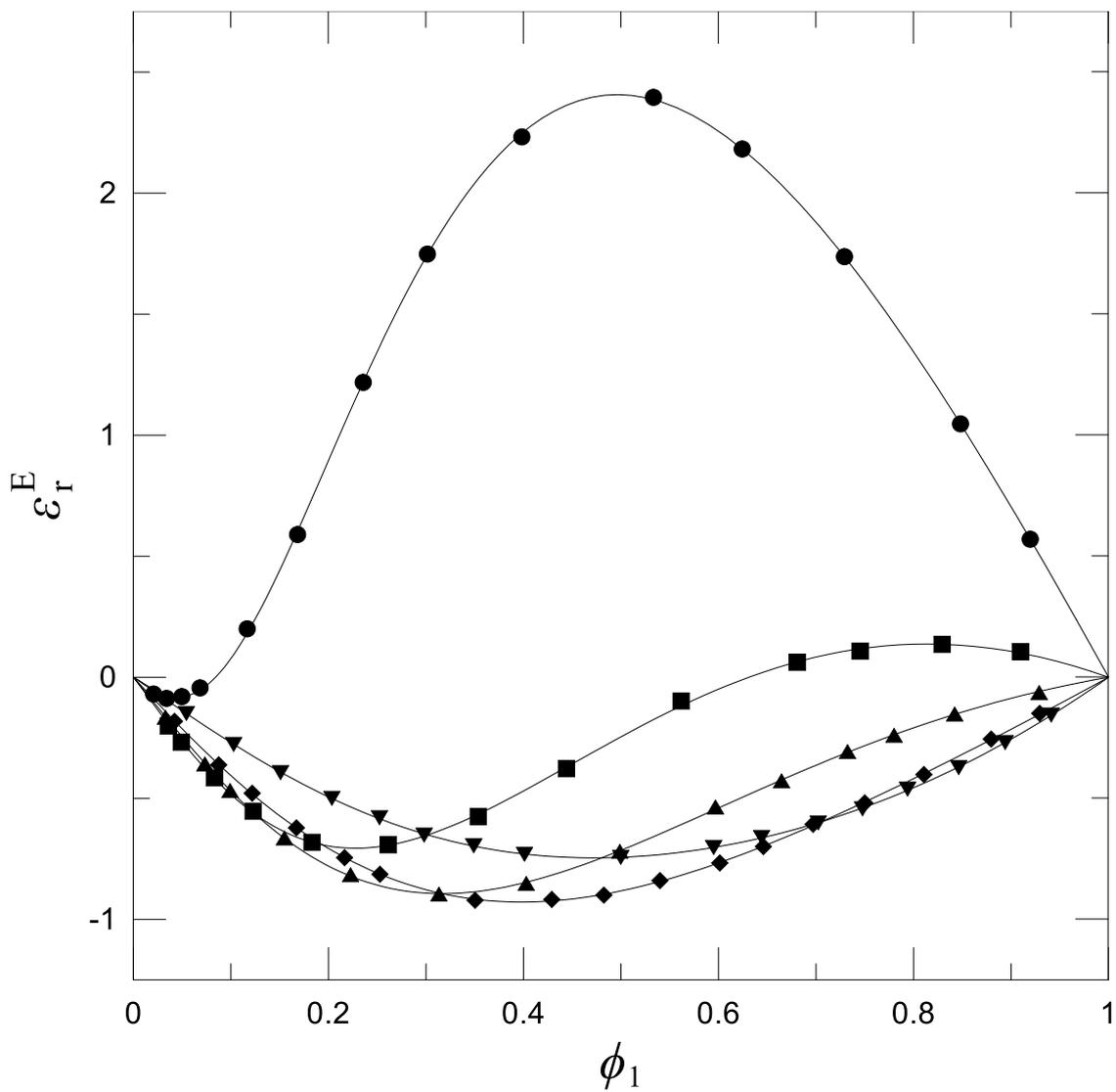

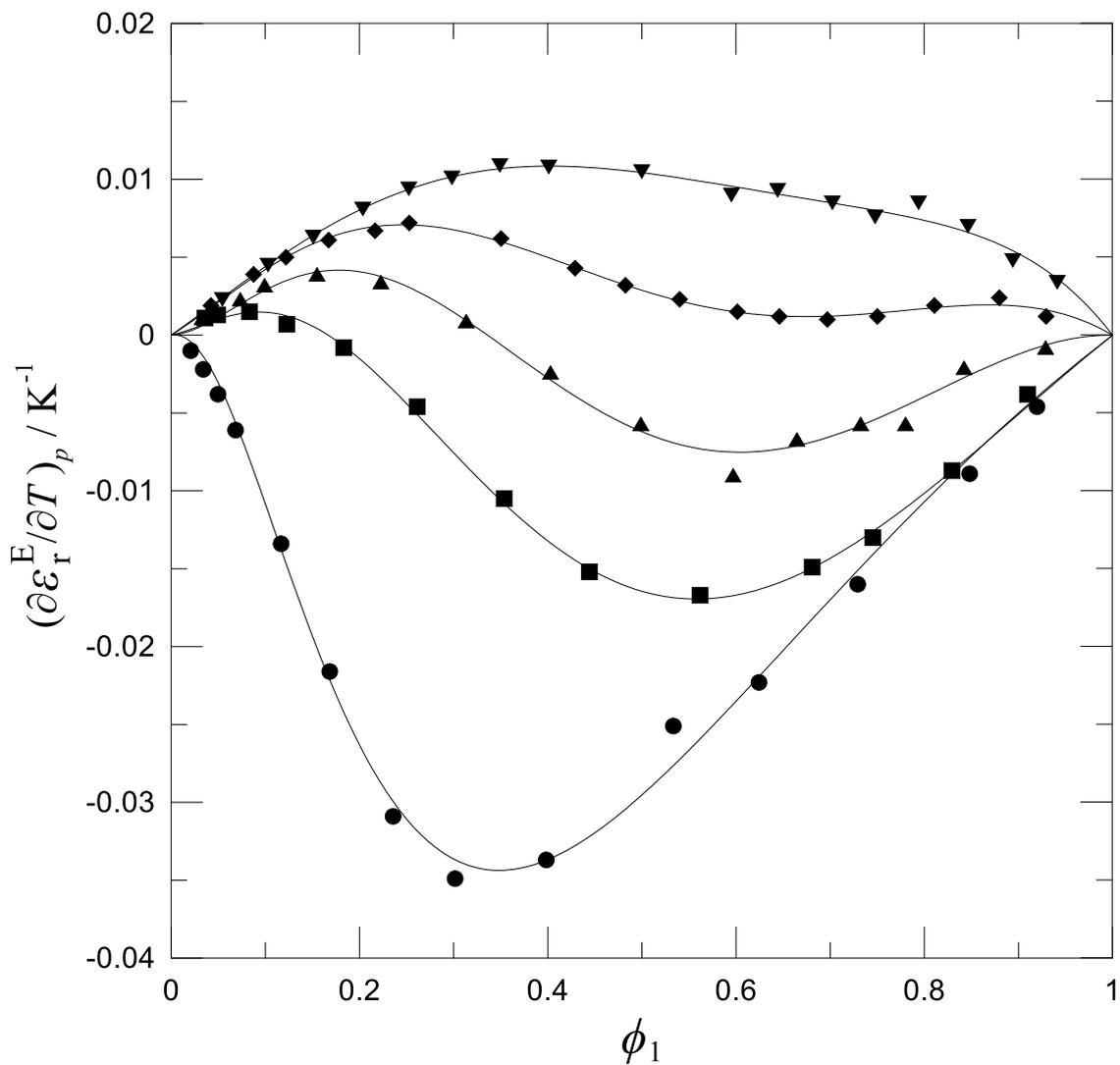

**Figure 2**: Derivative of the excess relative permittivity of 1-alkanol (1) + di-*n*-propylamine (2) systems at 0.1 MPa, 298.15 K and 1 MHz. Full symbols, experimental values (this work): (●), methanol; (■), 1-propanol; (▲), 1-butanol; (♦), 1-pentanol; (▼), 1-heptanol. Solid lines, calculations with equation (5) using the coefficients from Table 5.

**Figure 3**: Excess refractive index, $n_D^E$, of 1-alkanol (1) + di-*n*-propylamine (2) systems at 0.1 MPa, 298.15 K and 1 MHz. Full symbols, experimental values (this work): (●), methanol; (■), 1-propanol; (▲), 1-butanol; (♦), 1-pentanol; (▼), 1-heptanol. Solid lines, calculations with equation (5) using the coefficients from Table 5.

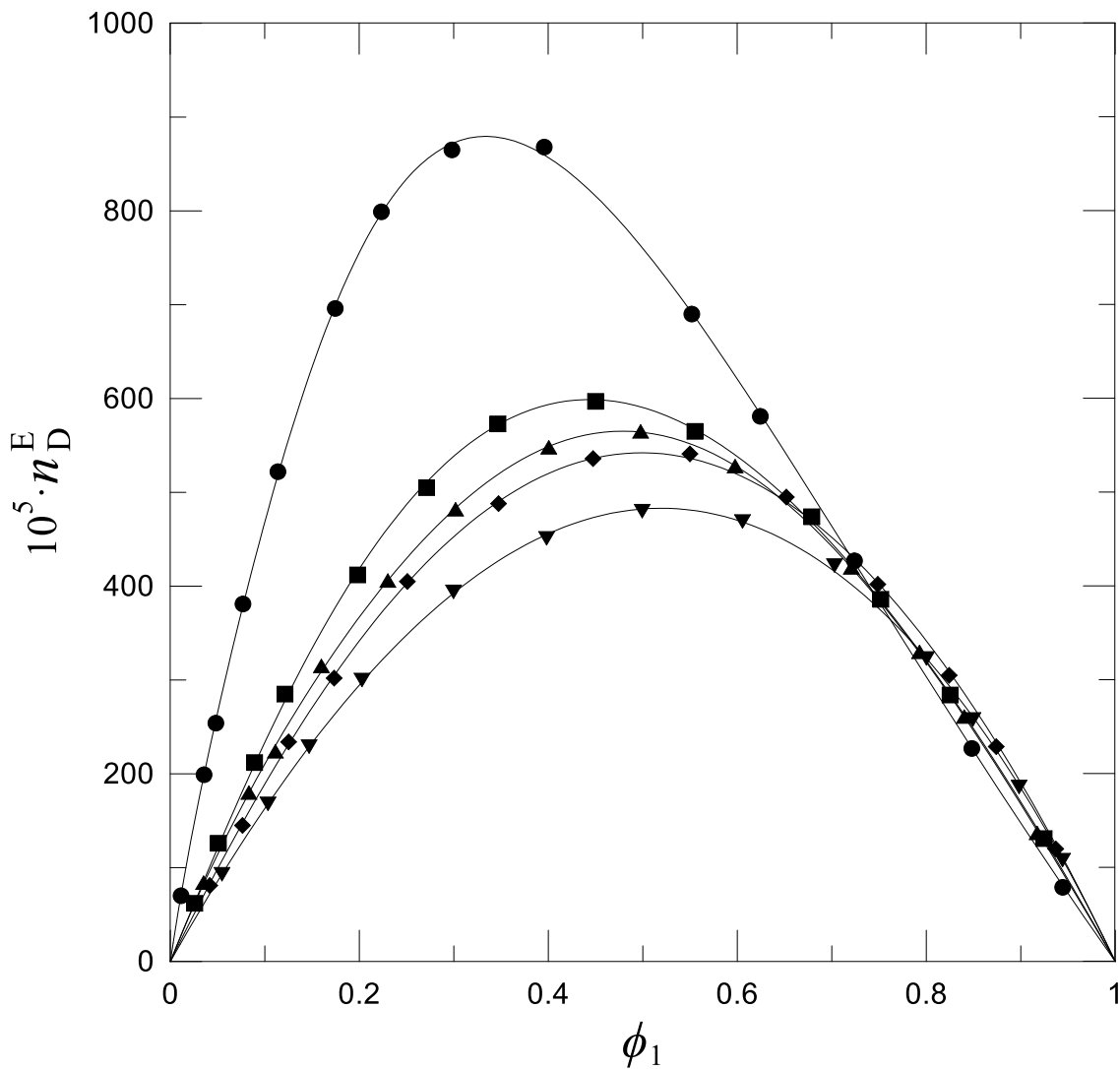

**Figure 4**: Excess relative permittivities at $\phi_1 = 0.5$ of 1-alkanol (1) + amine (2) or + heptane (2) systems as functions of the number of carbon atoms of the 1-alkanol, at 0.1 MPa, 298.15 K and 1 MHz: (●), *n*-hexylamine [23]; (▲), di-*n*-propylamine (this work); (♦), heptane [20, 43-45]; (■), di-*n*-propylether [51].

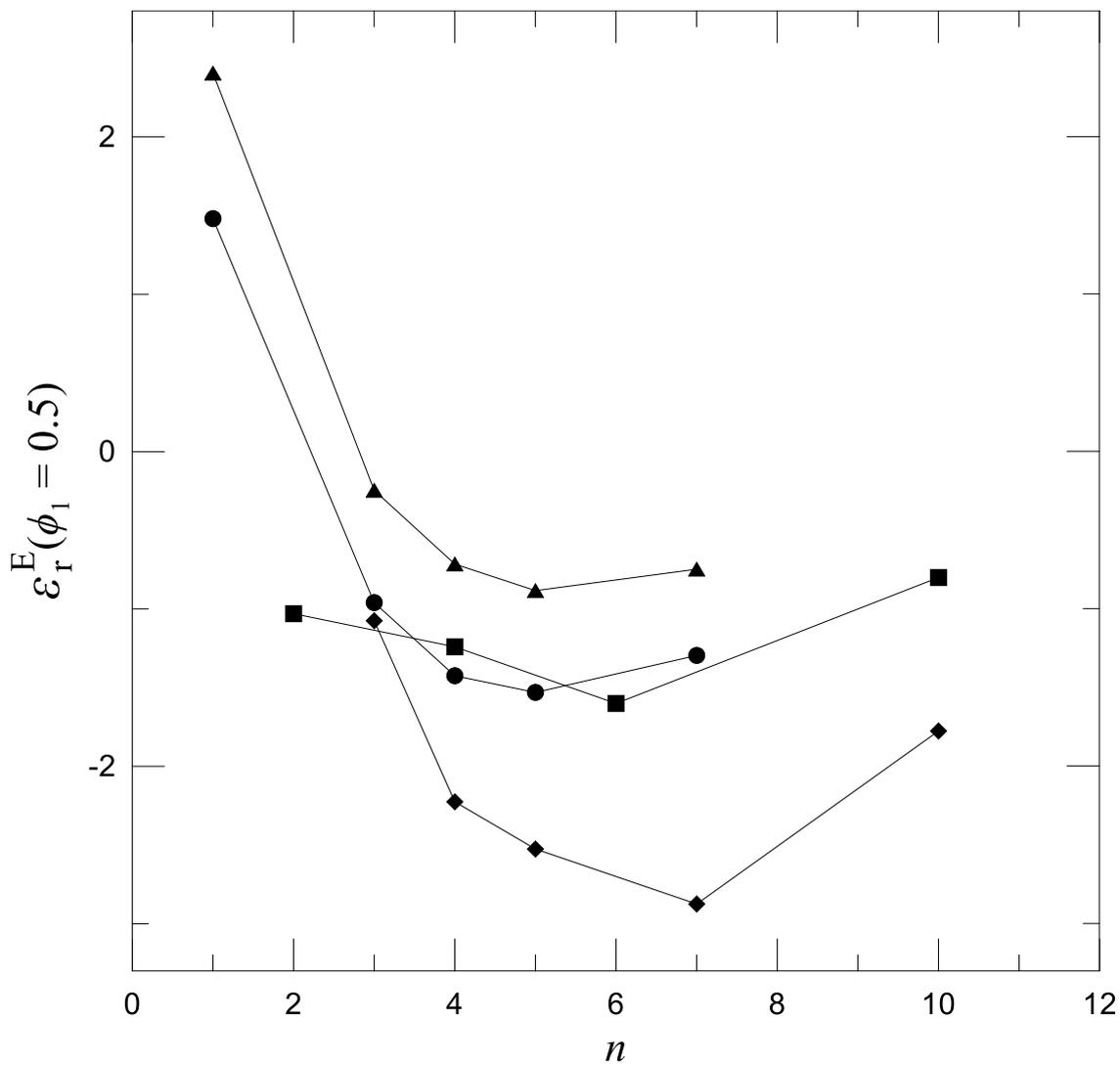

**Figure 5**: Kirkwood correlation factor, $g_K$, of 1-alkanol (1) + di-*n*-propylamine (2) systems at 0.1 MPa, 298.15 K and 1 MHz. Numbers in parentheses indicate the number of atoms of the 1-alkanol.

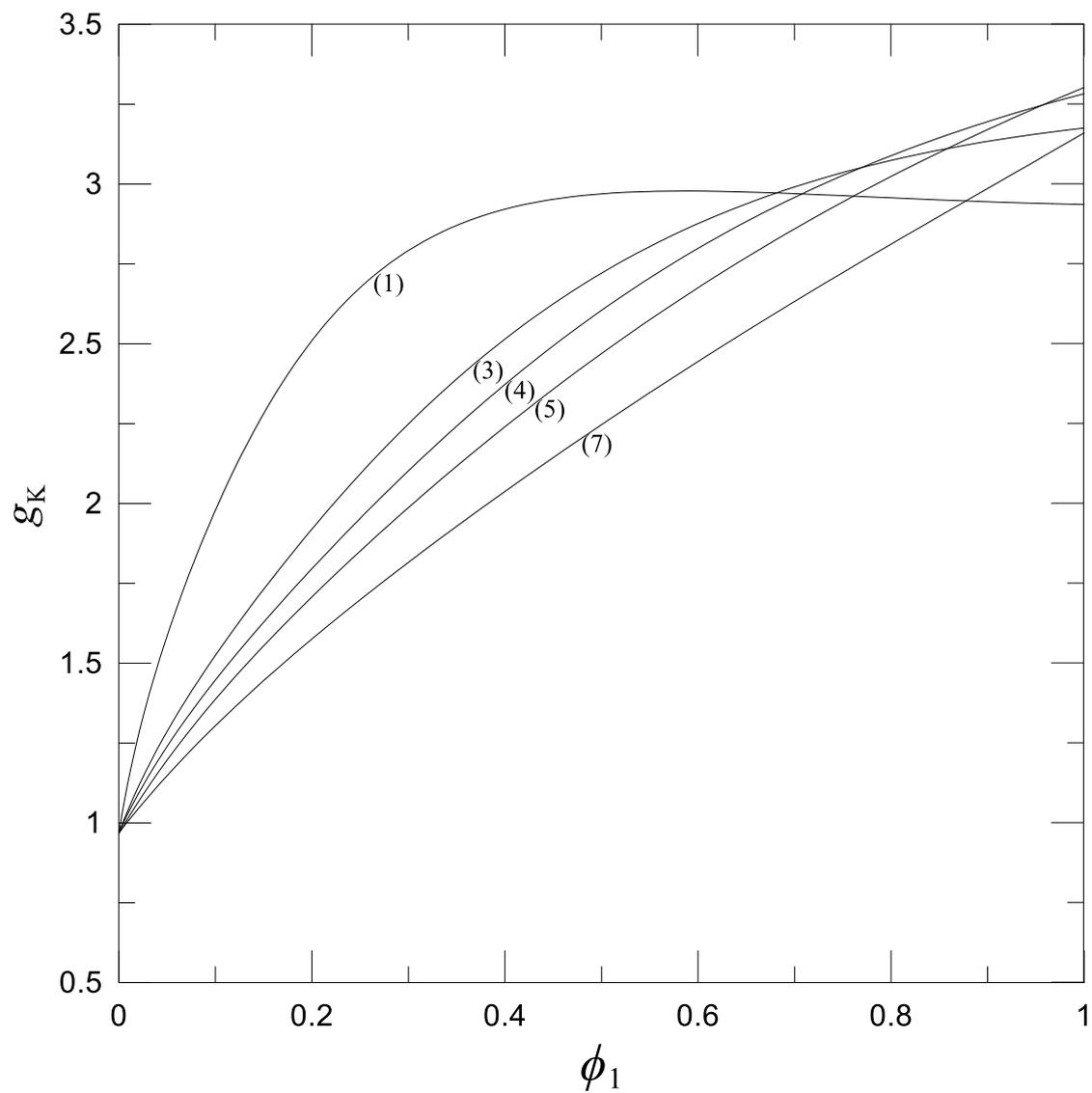

**Figure 6**: Excess Kirkwood correlation factor, $g_K^E$, of 1-alkanol (1) + di-*n*-propylamine (2) systems at 0.1 MPa, 298.15 K and 1 MHz. Numbers in parentheses indicate the number of atoms of the 1-alkanol.

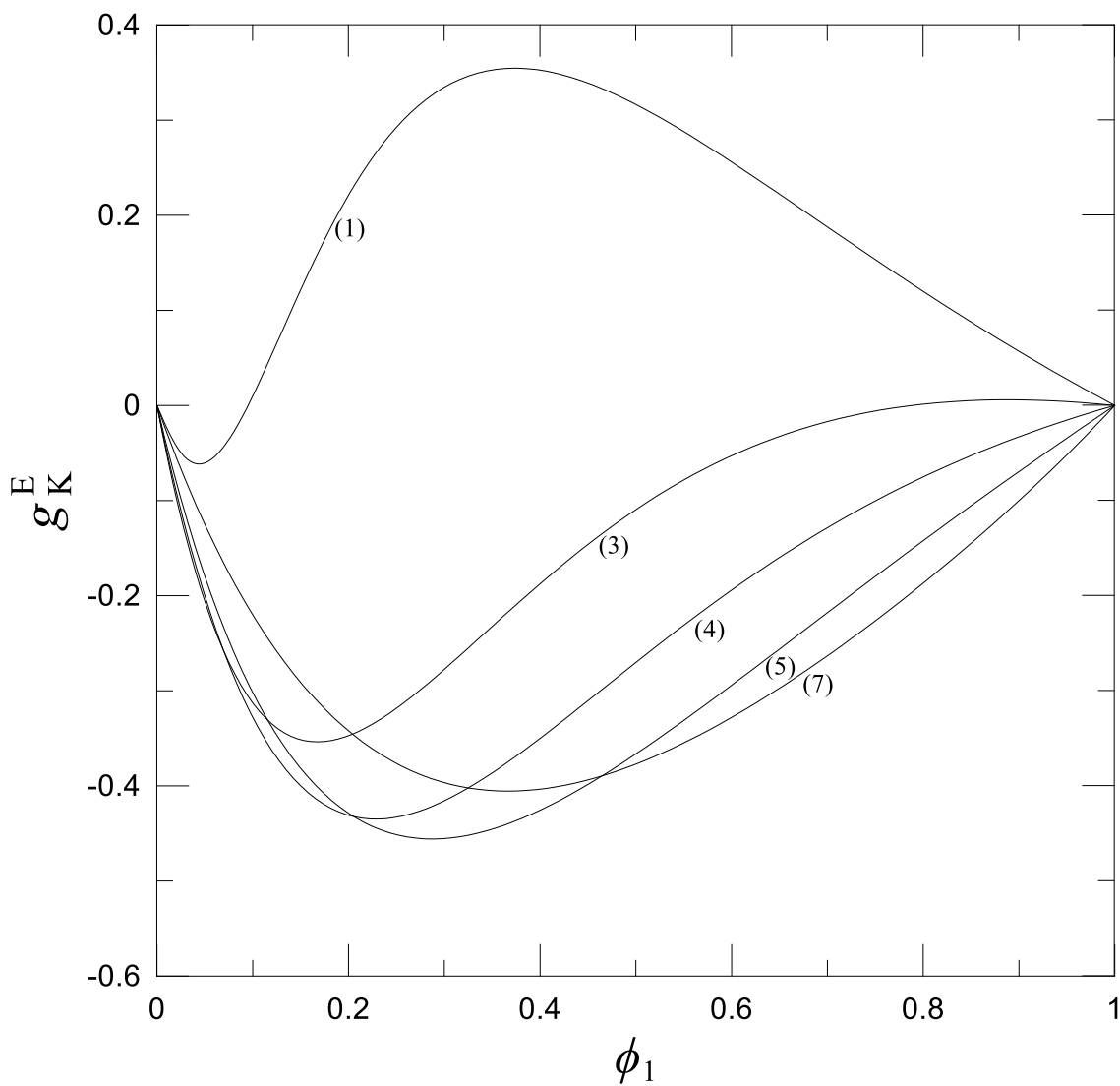

**Figure 7**: Excess Kirkwood correlation factors at $\phi_1 = 0.5$ of 1-alkanol (1) + amine (2) systems as functions of the number of carbon atoms of the 1-alkanol, at 0.1 MPa, 298.15 K and 1 MHz: (●), *n*-hexylamine [23]; (▲), di-*n*-propylamine (this work).

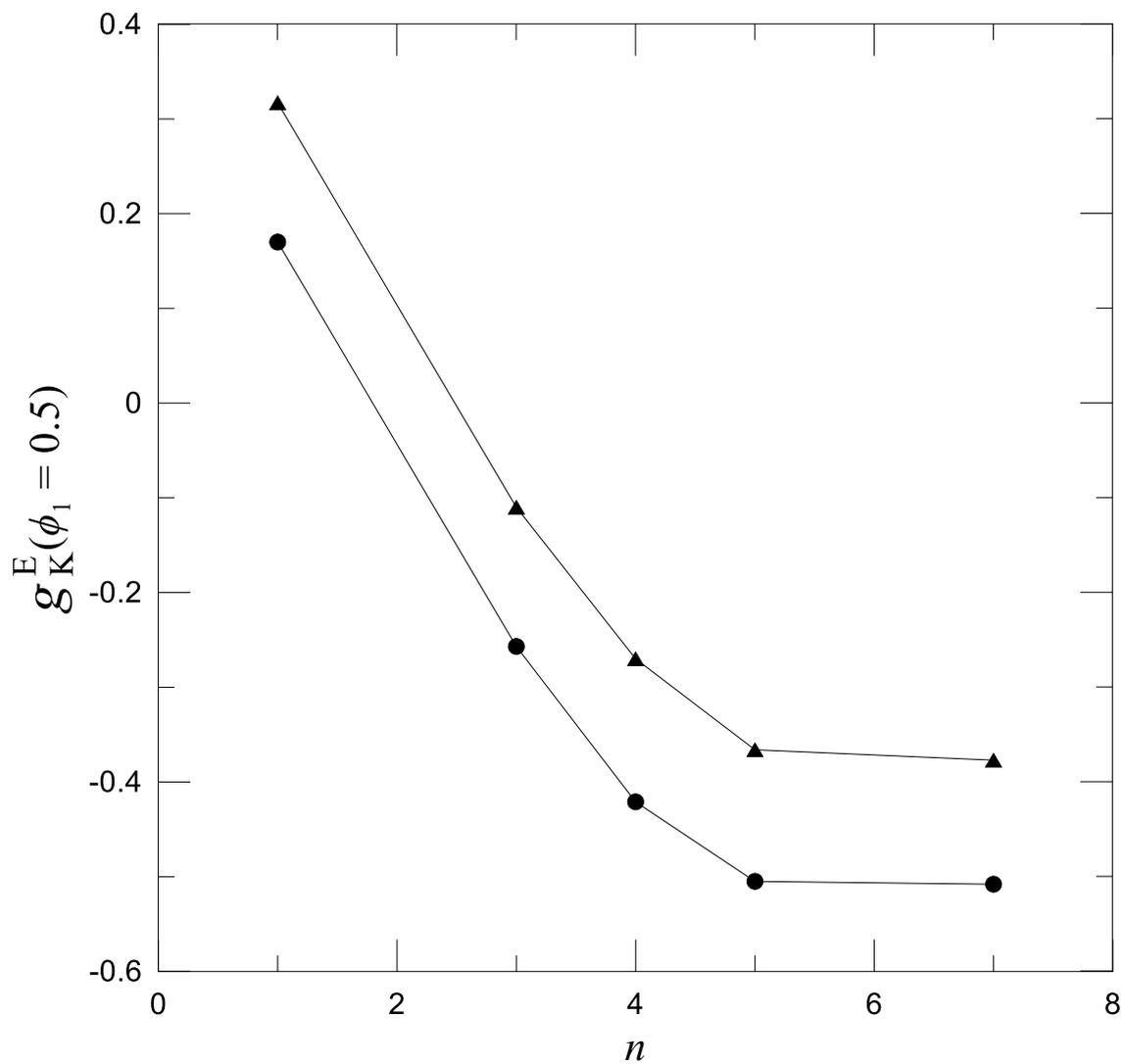

**Thermodynamics of mixtures with strongly negative deviations from Raoult's law. XVI. Permittivities and refractive indices for 1-alkanol + di-*n*-propylamine systems at (293.15-303.15) K. Application of the Kirkwood-Fröhlich model**

**Supplementary material**

Fernando Hevia, Ana Cobos, Juan Antonio González*, Isaías García de la Fuente, Luis Felipe Sanz

G.E.T.E.F., Departamento de Física Aplicada, Facultad de Ciencias, Universidad de Valladolid. Paseo de Belén, 7, 47011 Valladolid, Spain.

*e-mail: jagl@termo.uva.es; Tel: +34-983-423757

Table S1

Derivative of the excess relative permittivity of 1-alkanol (1) + di-*n*-propylamine (DPA) (2) systems at 0.1 MPa, 298.15 K and 1 MHz [a].

| $x_1$ | $\phi_1$ | $(\partial \varepsilon_r^E / \partial T)_p$ / K$^{-1}$ | $x_1$ | $\phi_1$ | $(\partial \varepsilon_r^E / \partial T)_p$ / K$^{-1}$ |
|---|---|---|---|---|---|
| | | methanol (1) + DPA (2) ; $T$/K = 298.15 | | | |
| 0.0664 | 0.0206 | -0.001 | 0.5940 | 0.3015 | -0.035 |
| 0.1066 | 0.0340 | -0.002 | 0.6918 | 0.3984 | -0.0337 |
| 0.1503 | 0.0496 | -0.004 | 0.7948 | 0.5333 | -0.0251 |
| 0.1990 | 0.0683 | -0.006 | 0.8492 | 0.6243 | -0.0223 |
| 0.3091 | 0.1166 | -0.013 | 0.9012 | 0.7291 | -0.016 |
| 0.4068 | 0.1683 | -0.022 | 0.9498 | 0.8481 | -0.0089 |
| 0.5110 | 0.2357 | -0.031 | 0.9749 | 0.9197 | -0.0046 |
| | | 1-propanol (1) + DPA (2) ; $T$/K = 298.15 | | | |
| 0.0638 | 0.0358 | 0.0011 | 0.5948 | 0.4442 | -0.0152 |
| 0.0866 | 0.0491 | 0.0013 | 0.7018 | 0.5617 | -0.0167 |
| 0.1427 | 0.0831 | 0.0015 | 0.7967 | 0.6809 | -0.0149 |
| 0.2045 | 0.1228 | 0.0007 | 0.8431 | 0.7453 | -0.0130 |
| 0.2917 | 0.1832 | -0.0008 | 0.8993 | 0.8294 | -0.0087 |
| 0.3939 | 0.2614 | -0.0046 | 0.9487 | 0.9097 | -0.0038 |
| 0.5012 | 0.3536 | -0.0105 | | | |
| | | 1-butanol (1) + DPA (2) ; $T$/K = 298.15 | | | |
| 0.0484 | 0.0328 | 0.0011 | 0.5991 | 0.4989 | -0.0057 |
| 0.1063 | 0.0734 | 0.0023 | 0.6896 | 0.5968 | -0.0090 |
| 0.1418 | 0.0992 | 0.0032 | 0.7484 | 0.6646 | -0.0067 |
| 0.2157 | 0.1549 | 0.0039 | 0.8041 | 0.7323 | -0.0057 |
| 0.3006 | 0.2226 | 0.0034 | 0.8418 | 0.7800 | -0.0057 |
| 0.4064 | 0.3133 | 0.0009 | 0.8890 | 0.8422 | -0.0021 |
| 0.5030 | 0.4028 | -0.0024 | 0.9514 | 0.9288 | -0.0008 |
| | | 1-pentanol (1) + DPA (2) ; $T$/K = 298.15 | | | |
| 0.0530 | 0.0422 | 0.0019 | 0.5985 | 0.5400 | 0.0023 |
| 0.1086 | 0.0875 | 0.0039 | 0.6570 | 0.6014 | 0.0015 |
| 0.1497 | 0.1218 | 0.0050 | 0.6985 | 0.6460 | 0.0012 |
| 0.2032 | 0.1672 | 0.0061 | 0.7450 | 0.6970 | 0.0010 |
| 0.2597 | 0.2165 | 0.0067 | 0.7921 | 0.7500 | 0.0012 |
| 0.3006 | 0.2529 | 0.0072 | 0.8447 | 0.8107 | 0.0019 |
| 0.4064 | 0.3503 | 0.0062 | 0.9027 | 0.8796 | 0.0024 |
| 0.4882 | 0.4290 | 0.0043 | 0.9435 | 0.9293 | 0.0012 |
| 0.5421 | 0.4825 | 0.0032 | | | |
| | | 1-heptanol (1) + DPA (2) ; $T$/K = 298.15 | | | |
| 0.0529 | 0.0543 | 0.0023 | 0.5883 | 0.5950 | 0.0090 |
| 0.1003 | 0.1028 | 0.0045 | 0.6378 | 0.6442 | 0.0093 |
| 0.1474 | 0.1509 | 0.0063 | 0.6965 | 0.7023 | 0.0085 |
| 0.1990 | 0.2035 | 0.0081 | 0.7424 | 0.7477 | 0.0076 |
| 0.2471 | 0.2523 | 0.0094 | 0.7892 | 0.7938 | 0.0085 |
| 0.2924 | 0.2982 | 0.0101 | 0.8426 | 0.8463 | 0.0070 |
| 0.3427 | 0.3490 | 0.0109 | 0.8913 | 0.8940 | 0.0048 |
| 0.3943 | 0.4010 | 0.0108 | 0.9395 | 0.9411 | 0.0034 |
| 0.4929 | 0.4998 | 0.0105 | | | |

[a] The standard uncertainties are: $u(T) = 0.02$ K; $u(p) = 1$ kPa; $u(\nu) = 20$ Hz; $u(x_1) = 0.0010$; $u(\phi_1) = 0.004$. The standard uncertainty is: $u\left[\left(\partial \varepsilon_r^E / \partial T\right)_p\right] = 0.0008$ K$^{-1}$.

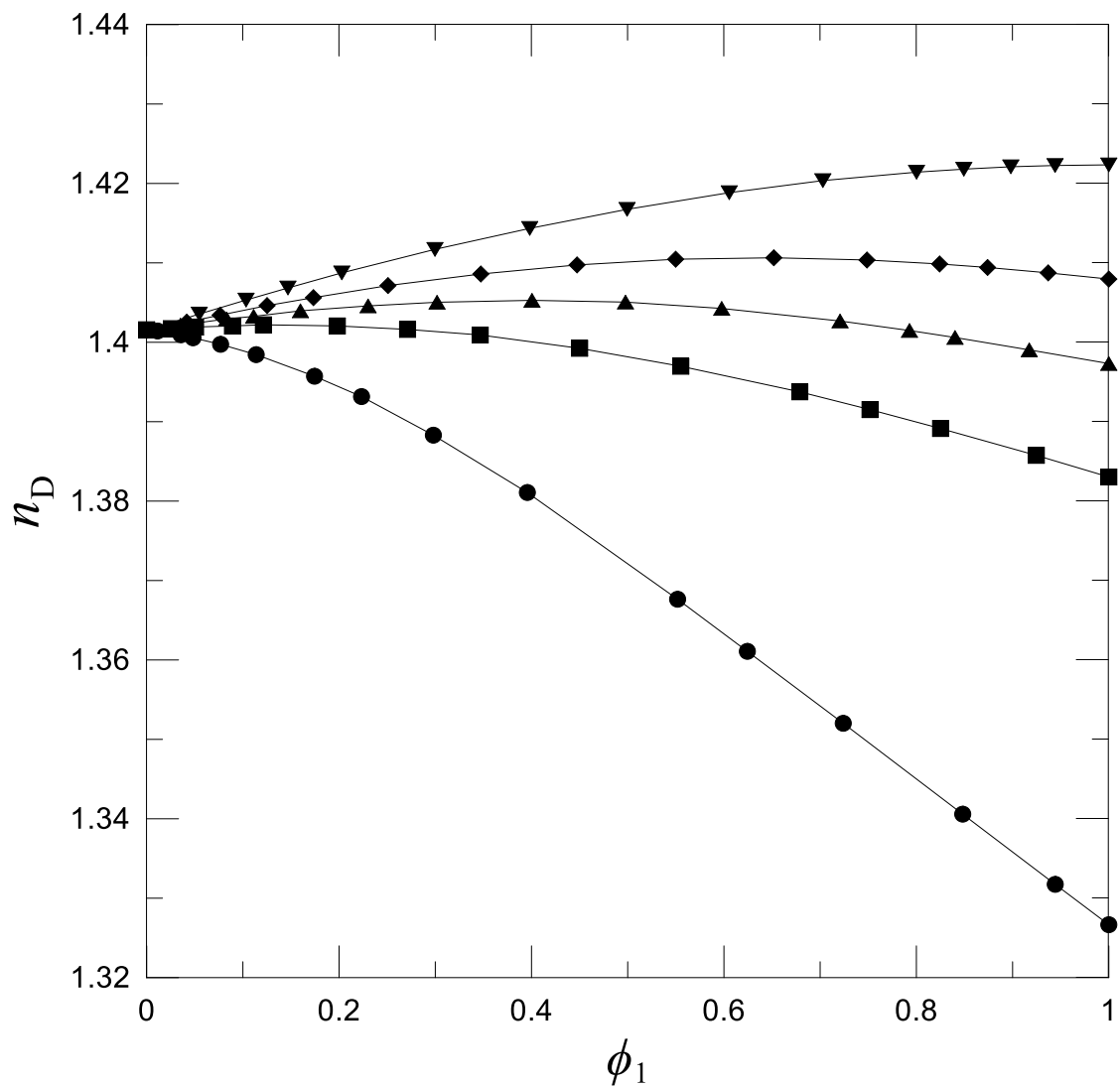

Figure S1

Refractive index at the sodium D line, $n_D$, of 1-alkanol (1) + di-n-propylamine (2) systems at 0.1 MPa and 298.15 K. Full symbols, experimental values (this work): (●), methanol; (■), 1-propanol; (▲), 1-butanol; (♦), 1-pentanol; (▼), 1-heptanol.

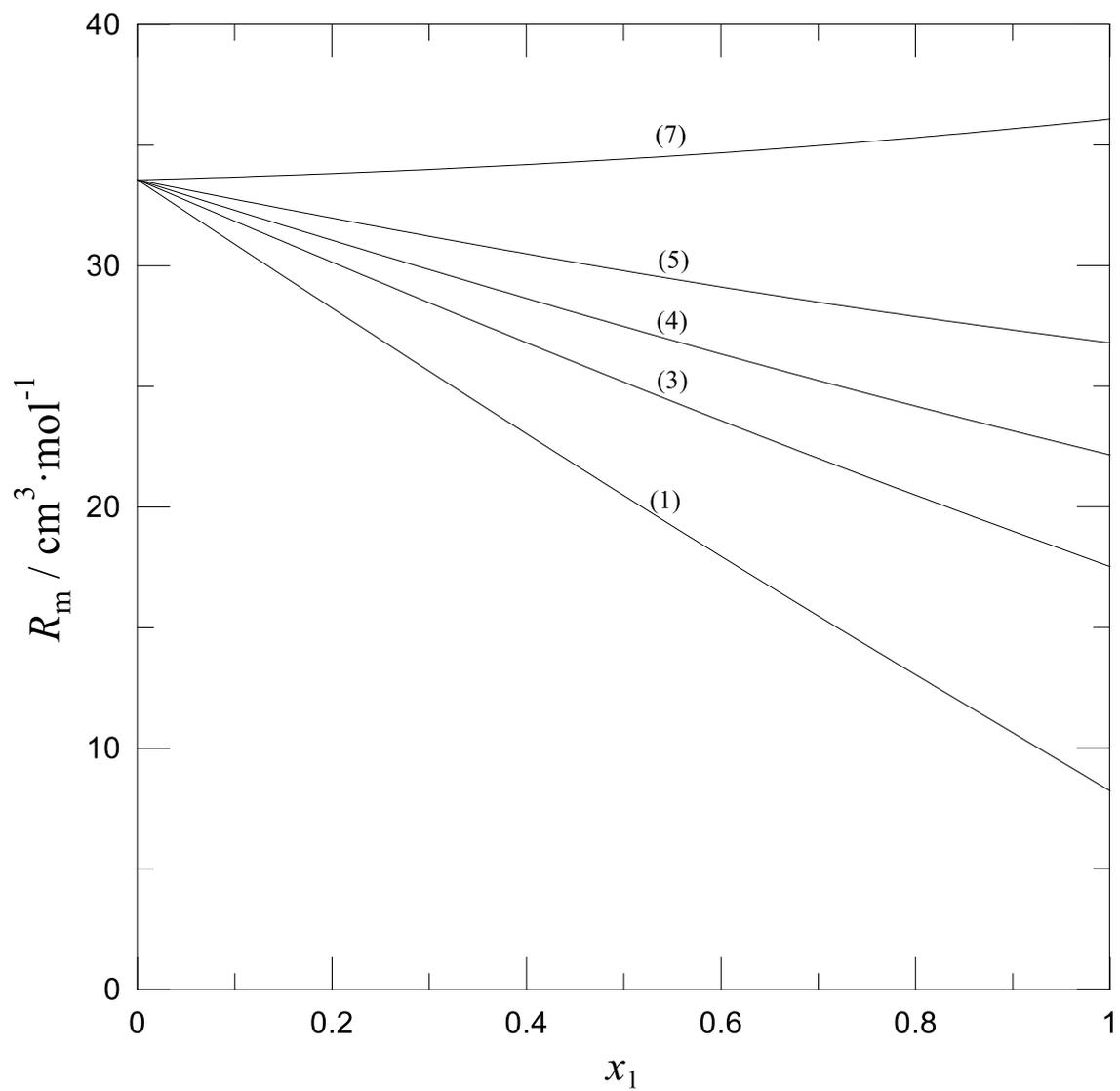

Figure S2

Molar refraction of 1-alkanol (1) + di-*n*-propylamine (2) systems at 0.1 MPa and 298.15 K. Numbers in parentheses indicate the number of atoms of the 1-alkanol.